\documentclass[journal,comsoc]{IEEEtran}
\usepackage[T1]{fontenc}

\newcommand{\subparagraph}{}
\usepackage{titlesec}
\titleformat*{\subsubsection}{\fontsize{9.5}{10}\bfseries}
\titleformat*{\subsection}{\fontsize{10}{11}\itshape}
\titleformat*{\paragraph}{\fontsize{9}{9.5}\bfseries}

\usepackage[hyphens]{url}

\usepackage{mathtools}

\usepackage{ragged2e}
\usepackage{blindtext, graphicx}
\usepackage{enumitem} 
\usepackage{hyphenat} 
\usepackage[tracking=true]{microtype} 
\SetTracking[no ligatures = f]{encoding = *,shape = sc}{120}
\usepackage{amsmath, amsthm, amssymb}
\usepackage{amsmath,amssymb,enumitem,mathtools}
\usepackage{xcolor}

\usepackage[most]{tcolorbox}
\newtcolorbox{highlighted}{colback=yellow,coltext=black,breakable}
\usepackage{lipsum}
\usepackage{float}
\usepackage{dblfloatfix}
\usepackage{soul}
\usepackage{xcolor}
\usepackage{graphicx}
\usepackage[export]{adjustbox}

\usepackage{color, soul}

\soulregister\cite7
\soulregister\ref7
\soulregister\pageref7

\ifCLASSOPTIONcompsoc
\usepackage[nocompress]{cite}
\else
\usepackage{cite}

\bstctlcite{IEEEexample:BSTcontrol}
\ifCLASSINFOpdf
\else
\fi
\ifCLASSOPTIONcompsoc
  \usepackage[caption=false,font=footnotesize,labelfont=sf,textfont=sf]{subfig}
\else

  \usepackage[caption=false,font=footnotesize]{subfig}
\fi
\begin{document}
\bstctlcite{IEEEexample:BSTcontrol}
	%
	\title{3RSeT: \underline{R}ead Disturbance \underline{R}ate \underline{R}eduction in STT-MRAM Caches by \underline{Se}lective \underline{T}ag Comparison}

\author{Elham~Cheshmikhani,~Hamed~Farbeh,~\IEEEmembership{Member,~IEEE}~and~Hossein~Asadi,~\IEEEmembership{Senior Member,~IEEE}
\thanks{E. Cheshmikhani and H. Asadi (corresponding author) are with the Department of Computer Engineering, Sharif University of Technology, Tehran 1115511365, Iran.\protect\\
			E-mail: elham.cheshmikhani@sharif.edu; asadi@sharif.edu}
\thanks{H. Farbeh is with the Department of Computer Engineering, Amirkabir University of Technology, Tehran  1591634311, Iran.\protect\\
			E-mail: farbeh@aut.ac.ir.}
\thanks{Manuscript received}}
	
	%
	%

	\markboth{IEEE Transactions on Computers,~Vol.~XX, No.~X, October~2019}%
	{Shell \MakeLowercase{\textit{et al.}}: Bare Demo of IEEEtran.cls for Computer Society Journals}
\maketitle
			\begin{abstract}
				
Recent development in memory technologies has introduced \textit{Spin-Transfer Torque Magnetic RAM} (STT-MRAM) as the most promising replacement for SRAMs in on-chip cache memories.
Besides its lower leakage power, higher density, immunity to radiation-induced particles, and non-volatility, an unintentional bit flip during read operation, referred to as \textit{read~disturbance} error, is a severe reliability challenge in STT-MRAM caches.
One major source of read disturbance error in STT-MRAM caches is simultaneous accesses to all tags for parallel comparison operation in a cache set, which has \textit{not} been addressed in previous work.
This paper first demonstrates that high read accesses to tag array extremely increase the read disturbance rate and then proposes a low-cost scheme, so-called \textit{\underline{R}ead~Disturbance \underline{R}ate~\underline{R}eduction~in~STT-MRAM~Caches~by~\underline{Se}lective~\underline{T}ag~Comparison} (3RSeT), to reduce the error rate by eliminating a significant portion of tag reads.
3RSeT proactively disables the tags that have no chance for hit, using low significant bits of the tags on each access request.
Our evaluations using gem5 full-system cycle-accurate simulator show that 3RSeT reduces the read disturbance rate in the tag array by 71.8\%, which results in 3.6x improvement in $Mean~Time~To~Failure$ (MTTF).
In addition, the energy consumption is reduced by 62.1\% without compromising performance and with less than 0.4\% area overhead.

			\end{abstract}
			

			\maketitle

			\section{Introduction}

\IEEEPARstart{O}{n-chip} cache memories play a decisive role in the system performance.
\textit{Static Random Access Memories} (SRAMs) have been the predominant technology in the cache memories for decades.
SRAM caches have faced several challenges by process technology downscaling in recent years. 
Nanoscale technology feature size has made SRAM caches highly vulnerable to soft errors, less scalable, and more power-hungry due to exacerbated leakage current~\cite{farbeh2016floating}.
Besides, process variations further degrade the efficiency of SRAM caches~\cite{farbeh2016floating, Eli-TR, wang2016memres}.
Therefore, an alternative technology is inevitable to overcome these challenges.

Among the emerging memory technologies, \textit{Spin-Transfer Torque Magnetic RAM} (STT-MRAM) is the most promising candidate for SRAM replacement in on-chip cache memories in the upcoming years~\cite{ZAZADTE, naeimi2013intel}. 
Non-volatility, high density, immunity to radiation-induced soft errors, and near-zero leakage power are among the main advantages of STT-MRAM~\cite{salkhordeh2019analytical, Chen2016,eli-aspdac}. 
Besides all of its advantages, the low reliability of STT-MRAM technology is a main challenge for system designers. 
Reading from and writing into the STT-MRAM cells are highly error-prone, which results in \textit{read~disturbance} and \textit{write~failure} errors, respectively~\cite{naeimi2013intel, mittal2017survey, Chintaluri-ESTCS2016, eli-date}.
Read disturbance is an unintentional flip of a cell during a read operation and write failure is an unsuccessful cell switching during a write operation~\cite{naeimi2013intel, choi2017nvm, mittal2017survey, Chintaluri-ESTCS2016, mittal2017addressing, eli-date}.

While the write failure rate is descending by smaller technology feature size, read current is not scaled well, which makes the read disturbance as the dominant error source in sub-32nm STT-MRAM caches~\cite{mittal2017addressing,bishnoi2014read, naeimi2013intel}. 
Read current, which is adjusted based on the physical characteristics of the memory device indicates the read disturbance probability of a single STT-MRAM cell per read access~\cite{bishnoi2014read, farbeh2018cache, 15-EDCC-zhao2011design, na2016read}.
On the other hand, read disturbance rate of STT-MRAM cell directly depends on its read intensity~\cite{eli-date, zhao2009high}.
On a cache access request (read and write), all tags in the target cache set are simultaneously read and compared with the tag part of the requested address, regardless of hit or miss event.
Reading from all $k$ tags in a k-way set-associative cache on each access makes the tag array highly vulnerable to read disturbance error.

Employing \textit{Error}-\textit{Detecting~and~Correcting~Codes} is the most conventional approach to protect on-chip caches against both transient and permanent errors.
Although data arrays in large L2/L3 caches are commonly protected using \textit{Error-Correcting~Codes} (ECCs), e.g., \textit{Single~Error~Correction}-\textit{Double~Error~Detection} (SEC-DED), the complexity of ECCs is barely affordable in high-speed tag arrays as well as L1 caches~\cite{naeimi2013intel, farbeh2016floating, intel, arm57, seyedzadeh2016leveraging}.
Hence, these memory structures remain unprotected or their protection is limited to simple \textit{Error-Detecting~Codes}, e.g., \textit{parity} coding~\cite{intel, arm57, farbeh2017raw, naeimi2013intel, farbeh2016floating}.

Some previous studies utilize ECCs to overcome read disturbance errors~\cite{eli-date, seyedzadeh2016leveraging, farbeh2016floating}.
$None$ of these schemes, however, targeted tag array in the cache and moreover, they are inapplicable or their overheads are unaffordable. 
Overwriting STT-MRAM cells after each read operation is another approach to tackle with read disturbance errors~\cite{wang2015selective, takemura2010highly}.
This approach substantially increases the energy consumption of highly-read tag array in addition to exacerbating the write failure rate.
Reducing the read current and designing more accurate sensing circuits to moderate its adverse effects on \textit{false~read} errors is an approach to decrease the read disturbance rate in STT-MRAM cells~\cite{15-EDCC-zhao2011design, na2016read, zhao2009high}.
The effectiveness of this approach is also limited due to unscalability of read current.
To the best of our knowledge, \textit{none} of the previous studies have addressed the vulnerability of tag~arrays to read disturbance in STT-MRAM caches.

In this paper, we propose \textit{\underline{R}ead~Disturbance \underline{R}ate~\underline{R}eduction~in~STT-MRAM~Caches~by~\underline{Se}lective~\underline{T}ag~Comparison} (3RSeT) scheme to reduce the rate of read disturbance errors in STT-MRAM tag arrays.
The key idea in 3RSeT is to decrease the number of reads from tag cells in  each access by eliminating a large fraction of unnecessary reads.
To this end, 3RSeT splits the tag comparison operation into two steps, instead of entirely reading all tags in one step.
In the first step, a few lower order bits of all tags are read and compared with the corresponding part of the requested address using a tiny comparator.
In the next step, the mismatched tag ways are disabled and the remaining bits of the other tags, if any, are read and compared with the corresponding bits of the requested address.
During these two steps, the operations in data array side proceed in the same manner as in the conventional cache architecture.
Since the bitwise similarity of tags in a set is likely to be reduced in lower order bits, the majority of tag ways is discarded from the second step. 
By eliminating a large fraction of read operations in tag array, 3RSeT significantly reduces the read disturbance rate.
This is while splitting the tag operation by 3RSeT has no performance penalty, because this operation overlaps the data array access and is completed beforehand.

We explore the efficiency of 3RSeT in reducing the error rate, which shows a strong dependency to the splitting point of tag array for partial comparison.
The larger number of bits contributing in the first step, the higher number of tags are discarded from the second step, while the higher error rate is imposed in the first step.
We conduct a comprehensive set of experiments on various values for splitting point and demonstrate that performing the partial comparison based on lower 4-bit tag part is the best selection for all workloads.

We evaluate 3RSeT using gem5 full-system cycle-accurate simulator~\cite{gem5} and compare it with the conventional tag comparison in STT-MRAM caches.
We Consider a 31-bit tag length\footnote{Considering a 48-bit address bus width as the typical configuration used in $Intel64$ and $AMD64$ processors family~\cite{address64,intel-xeon, AMD64,intel64}.} in a 1MByte 8-way set-associative L2 cache splitted into lower 4-bits part and higher 27-bit part in 3RSeT.
The evaluations show that 82.5\% of tag ways are disabled by 3RSeT in each access, on average.
By eliminating unnecessary reads from tag cells, 3RSeT reduces the read disturbance rate in tag array by an average of 71.8\%.
This reduction results in 3.6x increase in $Mean~Time~To~Failure$ (MTTF) of the cache.
The reduced number of read operations and comparisons results in 62.1\% energy saving in the tag array.
These significant improvements are achieved without increasing the cache access time.

The \textbf{main}~\textbf{contributions} of this paper are as follows:
			
			\begin{enumerate}
				\item This is the $first~study$ that addresses the read disturbance challenge in the tag array of STT-MRAM caches.
				Our study reveals that simultaneous accesses to all tags in the cache set for comparison operation makes the tag array the most vulnerable cache part. These observations demonstrate that the vulnerability of tag array to read disturbance error is by 32.1x higher than that of data array.
			 	 \item We propose the 3RSeT scheme to reduce read disturbance error rate in the tag array by eliminating a large fraction of tag reads.
			 	 This is achieved by $a)$ comparing the tags in the cache set based on some lower order bits of the tag in advance and $b)$ preventing reading from the remaining other upper tag part for those that mismatched in the first step.
			 	 \item We experimentally find an optimum point for the number of lower order bits of tags for partial comparison to guarantee the minimum read disturbance rate in the tag array.
			 	 Our observations illustrate that this optimum point is interestingly the same for all workloads with different behaviors and access patterns.
				Any value greater (smaller) than this point increases the error rate by dominating the contribution of lower (higher) order bits of the tags in the number of reads. 
			 	 \item By making a minor modification, we propose an efficient cache configuration for 3RSeT realization that beside error rate reduction significantly reduces the energy consumption in the tag array with no effect on cache latency and imposing a negligible area overhead. 
			 
			\end{enumerate}

The rest of this paper is organized as follows. Section II describes the preliminaries of STT-MRAM memory and its reliability challenges. 
The previous studies are discussed in Section III.
In Section IV, the motivation and the proposed scheme is presented.
In Section V, the simulation setup and results are given.
We discuss different aspects of the proposed scheme and its limitations in Section VI.
Finally, we conclude the paper in Section VII.

\section{Preliminaries}
			\label{sec:PRELIMINARIES}
			In this section, we explain the conventional structure of STT-MRAM cells and the mechanisms of reading/writing from/into a STT-MRAM cell.
			Then, we discuss the sources of errors in STT-MRAM cells and focus on the read disturbance error as the most important reliability concern in the tag array of STT-MRAM cache.

				\subsection{STT-MRAM Basics}
\begin{figure}[t]\vspace{-10pt}
				\centering
				\subfloat[]{\includegraphics[width=0.43\linewidth]{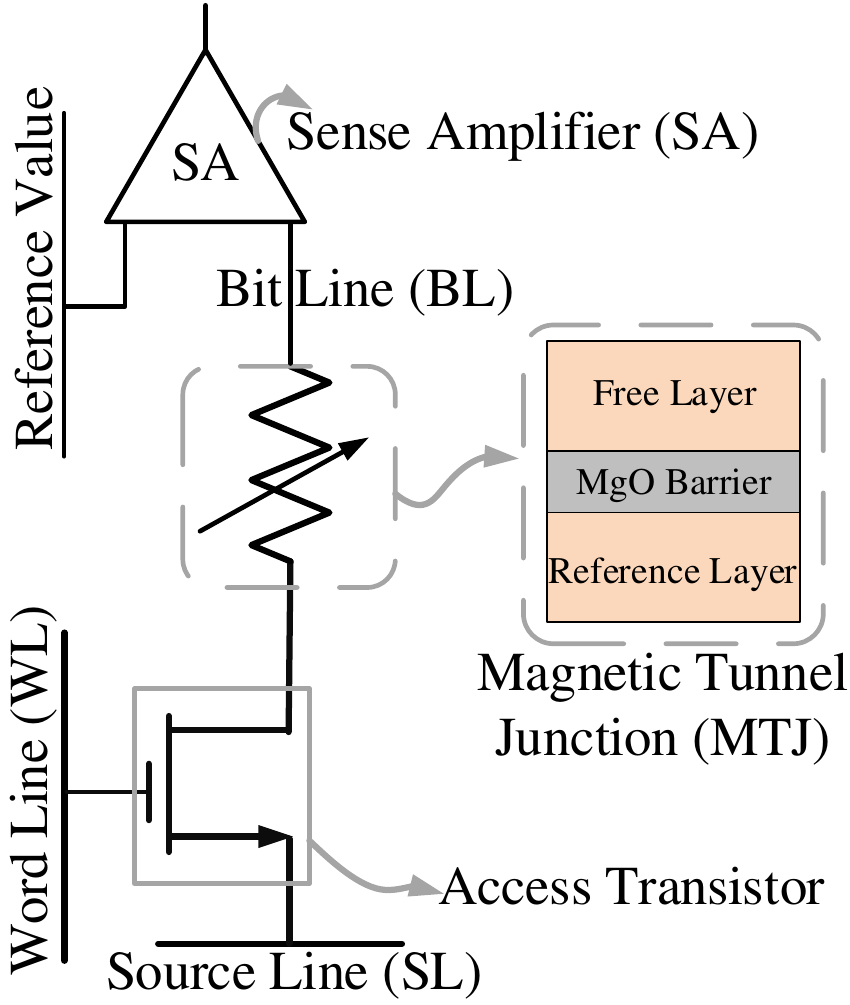}}
				\hspace{20pt}
				\subfloat[]{\includegraphics[width=0.44\linewidth]{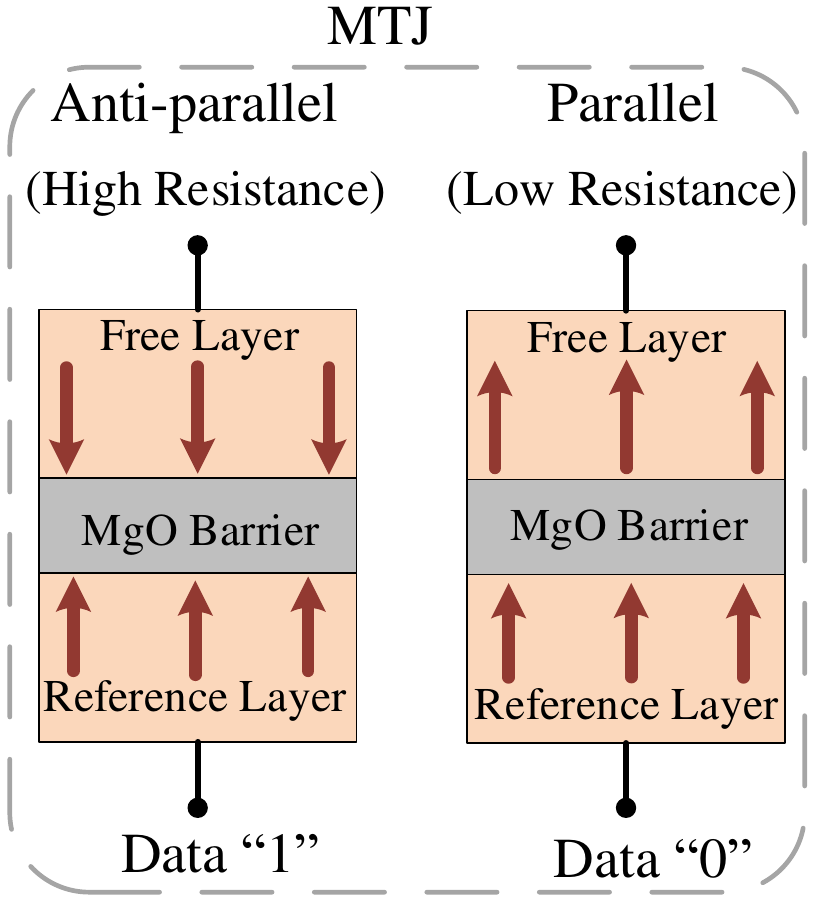}}
				\caption{STT-MRAM cell schematic: (a) 1T1J STT-MRAM cell structure and (b) MTJ low and high resistance states.}\vspace{-10pt}
				\label{fig:1}
			\end{figure}

STT-MRAM cell mechanism is based on resistance of a \textit{Magnetic~Tunnel~Junction} (MTJ) element, which determines the state of the cell~\cite{ahn2013selectively,farbeh2016floating}.
An NMOS access transistor controlled by the $Word~Line$ (WL) signal is used to connect/disconnect the $Bit~Line$ (BL) to the MTJ~\cite{wu2016temperature, farbeh2018cache,Eli-TC}.
MTJ consists of three layers, an oxide layer named \textit{oxide~barrier~layer} made of crystallized $Magnesium~oxide$ (MgO) sandwiched between two ferromagnetic layers, named \textit{Free~layer} and \textit{Reference~layer}~\cite{khvalkovskiy2013basic,apalkov2013spin, kang2015yield}.
The structure of this cell, known as 1T1MTJ STT-MRAM, is shown in Fig.~\ref{fig:1}(a). 
Magnetization direction of free layer can be changed by applying a write current and determines the state of stored data, while the magnetization direction of the reference layer is fixed.

STT-MRAM technology is based on the magnetic charge instead of electrical charge.
Two states of the MTJ is formed based on the free layer electrons spin direction while a spin-polarized current flow through it~\cite{vatajelu2015read, farbeh2018cache}.
This current makes the free layer spin direction parallel or anti-parallel with spin direction of the reference layer~\cite{wang2018adaptive}.
The parallelism causes a low resistance in MTJ interpreted as value `0', while `1' is because of high resistance due to anti-parallelism.
The MTJ states and its logic values are depicted in Fig.~\ref{fig:1}(b).


			As mentioned, the resistance of MTJ ($high~resistance$ or $low~resistance$) shows the value in a STT-MRAM cell (`1' or `0').
			This resistance value should be sensed to read the cell content.
Reading a cell needs a small current flowing from BL to $Source~Line$ (SL) or vice-versa for a predetermined pulse width~\cite{Pajouhi2016JETC, wang2018adaptive}.
In this case, WL is first set to turn on the access transistor.
Then, $I_{read}$ (the small read current) is applied to the STT-MRAM cell.
By applying $I_{read}$ to an STT-MRAM cell, a voltage is generated between the BL and SL. 
To read out the MTJ resistance state, the BL voltage should be compared with the reference voltage~\cite{mittal2017survey, 15-EDCC-zhao2011design}.	
If the sensed value is higher (lower) than the reference voltage, it means that the resistance of MTJ is low (high) and the cell contains `0' (`1') value.

To write a value into a cell, the MTJ resistance should be changed, which makes this operation more complicated.			
MTJ resistance changes if the spin direction of the electrons in the free layer flips. 
To this end, a write current ($I_{write}$) is applied to the bit line or source line to write `1' or `0', respectively.
Based on the direction of the applied current, the spin of free layer orients in the same or opposite direction of the reference layer magnetic field, which causes to flow a spin-polarized current. 
When the amount of this spin-polarized current exceeds a threshold value, the magnetic field direction of the free layer switches~\cite{Eli-TC, apalkov2013spin}. 
This is the time when the MTJ content flips and a value is written into the cell.
By switching the magnetic field direction from parallel to anti-parallel (or vice versa), electrons flow from the free layer to the reference layer (or vice versa)~\cite{mittal2017survey, 15-EDCC-zhao2011design, farbeh2018cache,kang2014variation}. 

\subsection{STT-MRAM Reliability} 

		Reliability is a major concern of STT-MRAM technology for commercialization. 
To address the reliability concern in STT-MRAM, it is required first to investigate the sources of errors in this technology.
Three main sources of errors in a STT-MRAM cell are $read~disturbance$, $write~failure$, and $retention~failure$.


Read disturbance error occurs when the flow of the read current through MTJ layers changes the electrons spin direction, resulting in an undesirable write operation. 
The read disturbance probability of an STT-MRAM cell during a read operation is calculated according to (\ref{eq:1})~\cite{Jiang-ASPDAC-16, bishnoi2014read, naeimi2013intel}.
\begin{equation}
			\begin{multlined}
			\label{eq:1}
			\shoveright[2cm]{P_{Read-Disturbance}=1- \exp(\frac{-t_{read}}{\tau\times\exp(\frac{\Delta(1-I_{read})}{I_{C_0}})})} 
			\end{multlined}
			\end{equation}
			where, \textit{$\tau$} is attempt period and assumed to be 1ns, \textit{I$_{read}$} is read current, \textit{I$_{C0}$} is critical switching current, which is needed to write in 0$^{\circ}$K, \textit{t$_{read}$} is the read pulse width, and $\Delta$ is $thermal~stability~factor$~\cite{vatajelu2015read, Eli-TC}.
The thermal stability factor of a cell is calculated according to (\ref{eq:4})~\cite{naeimi2013intel, vatajelu2015read}.\vspace{-8pt}
		
			\begin{flalign}
			\label{eq:4}
			\resizebox{.18\linewidth}{!}{$\Delta = \frac{E_{b}}{ k_{B} T}$}
			\end{flalign}
			where, \textit{E$_b$} is barrier energy, \textit{k$_B$} is $Boltzmann$ constant, and \textit{T} is temperature in Kelvin.

A write~failure occurs when the content of a cell is not switched by applying write current during the write operation.
This probability is calculated according to (\ref{eq:2})~\cite{naeimi2013intel,EDCC, eli-date}.
\begin{equation}
			\begin{multlined}
			\label{eq:2}
			 P_{Write-Failure} = \exp( -t_{write}\times \\
			\shoveright[1cm]{\frac{2 \times \mu_{\beta}\times p\times(I_{write}-I_{C_0})}{c+\log_{e}(\pi^2\times\Delta/4)\times (e\times m\times (1+p^2))})} 
			\end{multlined}
			\end{equation}		
			where, \textit{I$_{write}$} is write current, \textit{c} is $Euler$ constant, \textit{e} is electron charge, \textit{m} is magnetic momentum of the free layer, \textit{p} is tunneling spin polarization, \textit{$\mu$$_{\beta}$} is $Bohr$ magneton, \textit{t$_{write}$} is write pulse duration, and $\Delta$ is thermal stability factor.

Retention~failure occurs when the cell content flips stochastically, while the cell is idle (a cell that is not read nor written)~\cite{naeimi2013intel,Pajouhi2016JETC,Na-TCAS-II-16,Ran2016JSA, Eli-TR}. 
The occurrence probability of a retention failure for a STT-MRAM cell is according to (\ref{eq:3}).
				\begin{flalign}
			\label{eq:3}
			\resizebox{.8\linewidth}{!}{$ P_{Retention-Failure} = 1- \exp({-t }\times {exp(-\Delta)})$}
			\end{flalign}
			where, $t$ is the cell idle time and $\Delta$ is thermal~stability~factor of a STT-MRAM cell.

According to the mentioned formulas, the rate of all the error types in a STT-MRAM cell depends strongly on its physical- and circuit-level parameters, e.g., $\Delta$, $I_{read}$, $t_{read}$, $I_{write}$, and $t_{write}$. 
The retention time reported for a STT-MRAM cell is typically 10 years, which is large enough for making the retention failure rate ignorable in frequently accessed on-chip caches.
The rates reported for read disturbance and write failure reside within $10^{-8}$$\sim$$10^{-11}$ interval per STT-MRAM cell per access.
This value is large enough to make the read disturbance and write failure a severe reliability challenge in frequently read/written on-chip caches.
In a cache accessed for millions of times per second, experiencing several errors per unit of time is highly probable.

The tag array part in the cache is read for both data read and write requests in the $cache~hit$ and $cache~miss$ occurrences.
On the other hand, in a set-associative cache, which is the common cache organization in modern processors, all tag ways in a set are simultaneously read per request for a single data block.
Therefore, unlike data array, the number of read operations is significantly higher than write operations in the tag array, which makes the read disturbance as the predominant source of error in the tag array. 
Meanwhile, technology scaling further exacerbates the contribution of read disturbance in the total tag array error rate.
This is because the scaling of read current is significantly lower than that of write current by shrinking of technology feature size, which leads to narrow the gap between the read and write current~\cite{naeimi2013intel}.



\begin{figure*}[t]
				\centering
				\includegraphics[width=0.88\linewidth]{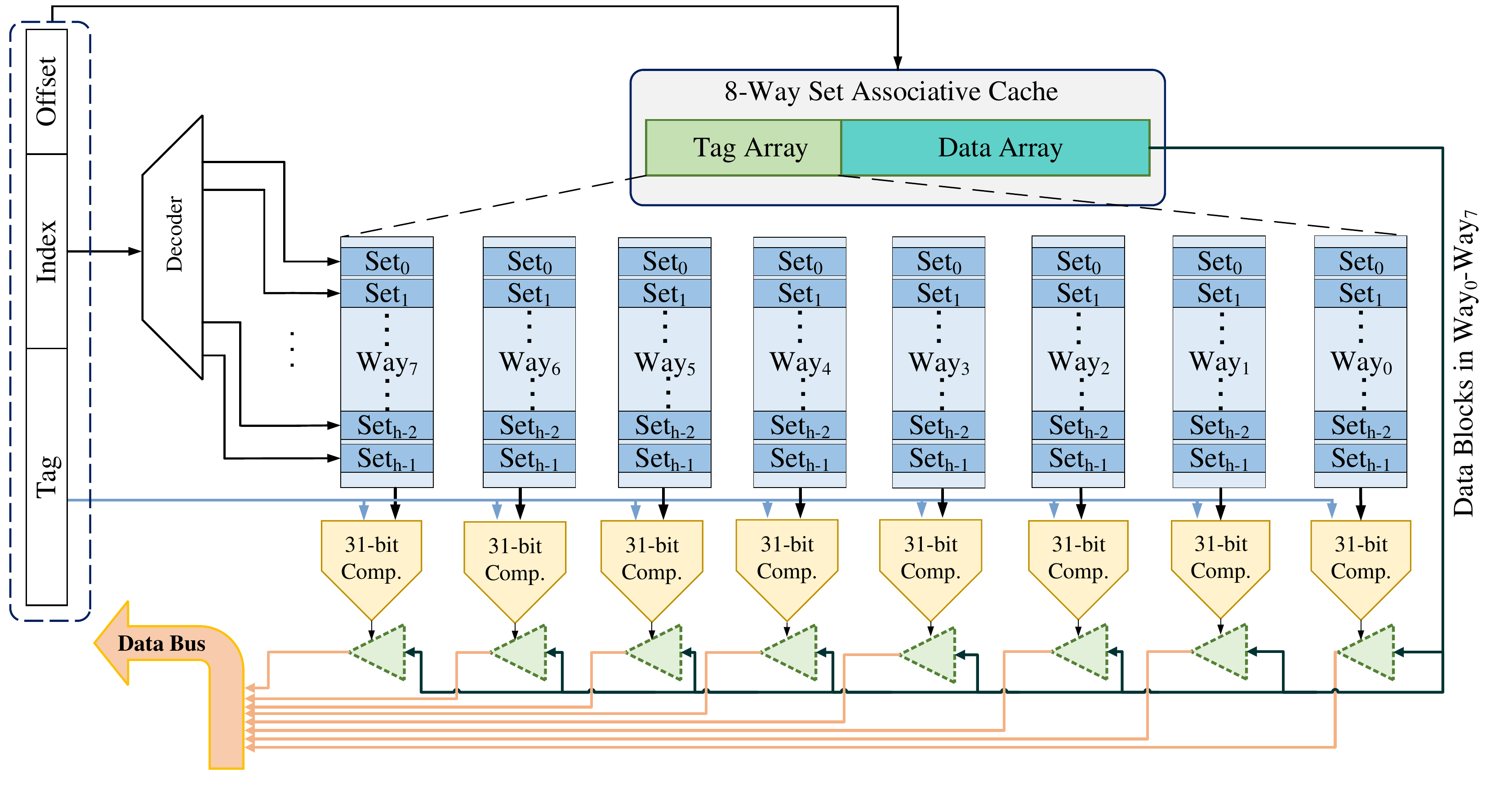}\vspace{-12pt}
				\caption{Structure of tag array in conventional cache configuration.}\vspace{-12pt}
				\label{fig:2}
\end{figure*}

\section{Related Work}

Several studies have addressed the challenges of STT-MRAM technology, which is widely used in both data and tag arrays of cache memories~\cite{ghaemi2015lated, priya2019cache, ghaemi2019sleepy,15-EDCC-zhao2011design, na2016read, zhao2009high,zhang2017shielding, takemura2010highly, wang2015selective,mittal2017addressing, aliagha2019react,eli-date, seyedzadeh2016leveraging}.
While overcomming read disturbance error in data array is the concern of several studies, this reliability threat has $never$ been addressed in tag array.
In the following subsections, we first discuss the previous work on read disturbance of data array of STT-MRAM caches to show that these schemes are not applicable/affordable in the tag array.
Then, we investigate the studies in traditional SRAM-based tag array aimed to reduce tag accesses for energy saving, because of their applicability in emerging STT-MRAM tag array and their effect on read disturbance rate.




\subsection{Read Disturbance Reduction in STT-MRAM-based Data Array}
Some of previous studies for read disturbance rate reduction are based on adjusting the circuit parameters.
They decrease the read current and/or read pulse width to mitigate read disturbance rate~\cite{15-EDCC-zhao2011design, na2016read, zhao2009high}.
This approach, however, increases the rate of $false~read$ error, which is the inability of correctly detecting the cell content.

Another approach on data array, named $Read$-$restore$ scheme~\cite{zhang2017shielding, takemura2010highly, wang2015selective}, overwrites the cells content after each read access.
This scheme is only able to correct the read disturbances that flip the cell content while the cell is correctly read in the ongoing access.
However, the stochastically occurring read disturbance may flip the cell in the initial phase of the read pulse and affect the ongoing read access as well.
In addition, the overhead of the energy-hungry and long-lasting write operation into cells, which are performed after each read from parallel-accessed tag array, is not affordable.

Another scheme uses data compression to reduce the number of bits written to the data blocks~\cite{mittal2017addressing, aliagha2019react}. 
By reconstructing the original data from its compressed form, this scheme omits a fraction of the read accesses from STT-MRAM cells.
Compressing and reconstructing data requires complicated circuitry and imposes performance overhead.
Some other studies utilize $Error$-${Correcting~Codes}$ (ECCs) to tolerate read disturbance errors~\cite{eli-date, seyedzadeh2016leveraging}.
Although ECCs are widely used to overcome read disturbance error in the data array, they severely degrade the cache performance when employed in the tag array.
Considering an ECC-equipped data array, the study in \cite{eli-date} demonstrates that frequent read accesses to cache blocks cause read disturbance accumulation in data blocks and proposes $REAP$-$Cache$ architecture to eliminate this accumulation.


\subsection{Read Access Reduction in SRAM-based Tag Array}
Considering the exacerbated read disturbance rate in the tag array due to its high read demand, this paper redesigns the tag structure to minimize the number of required reads.
To the best of our knowledge, $none$ of the previous studies have addressed the effects of high read accesses in tag array on the reliability of STT-MRAM caches.
However, there are some studies on traditional SRAM caches that try to reduce the cache energy consumption by eliminating a subset of extra read accesses in data and/or tag arrays.
Therefore, we studied the well-known previous schemes, which reduces read operations in tag part of cache.

$Way~prediction$ scheme tries to reduce the energy consumption imposed by parallel accesses in set-associative caches by accessing a cache way with the highest chance instead of accessing all cache ways~\cite{calder1996predictive,inoue1999way,sleiman2012embedded}. 
On a misprediction, this scheme accesses all cache ways to find a match.
Way prediction can be employed in only a) data array or b) both tag and data arrays.
In the former, instead of accessing all tag and data ways in a set, all tags and a single predicted data way is read.
On a misprediction, all data ways are accessed in parallel without requiring to repeat the tag read and compare operations.
In this manner, although the performance penalty of misprediction is minimal, the number of tag read accesses is not affected and no read disturbance reduction is achieved.
In the latter, a single predicted tag way and its corresponding data way are accessed, and on a misprediction, all tag and data ways are accessed in parallel.
Beside significant energy saving, this scheme can significantly reduce the tag read disturbance rate on a correct prediction.
However, way prediction is mainly effective in L1 caches and its misprediction rate in lower level caches is so high that almost no reliability improvement or energy saving is achieved, based on our experiments given in Section V.

As another scheme, $way~halting~cache$ inserts a narrow-width fully associative tag named $halt~tag$ beside each tag way to store some lowest-order bits of the corresponding tag~\cite{zhang2005way}.
Prior to accessing tag and data ways, all halt tag lines are accessed and compared with the corresponding lower order bits of the incoming tag.
Halt tag reduces read accesses by disabling the ways mismatched in the first phase.
Besides requiring extra memory cells, halt tag is only applicable to small L1 caches since the complexity, energy consumption, and delay of a fully-associative memory in lower level caches with thousands of lines are extremely higher than its benefits.

As a combination of two mentioned schemes, $way~halting~prediction$~\cite{mallya2015way} compares some lower order bits of incoming tag with the content of the fully associative halt tag to disable some cache ways and then access a single predicted way among the remaining ways.
Despite of its improvements in small L1 caches, the shortcomings of both way prediction and way halting cache make the way halting prediction scheme inapplicable in lower level caches.


The last group of the schemes that targeted read access reduction try to deactivate some data ways via partial tag comparison~\cite{kim2019segmented,peng2006low,min2004partial}.
A narrow-width tag way containing some lower order bits of the tag is inserted alongside of each tag way.
On each cache access request, a partial tag comparison is performed between the lower order bits of incoming tag and all narrow-width tags in a set; simultaneously, all tag and data ways are accessed similar to the conventional cache operation.
The output of each partial tag comparison is the enable signal of its corresponding data way sense amplifier unit.
Due to the fact that a large amount of energy is consumed by data array sense amplifiers, disabling some of them can result in a significant energy saving.
However, no reduction is provided by this scheme in the tag read access. 

\textbf{To~summarize,} the previous studies $only$ have addressed the read disturbance in data part of STT-MRAM caches.
The existing read disturbance reduction schemes on data array such as $a)$ changing the circuit parameters, $b)$ overwriting the memory content, $c)$ data coding and compression, and $d)$ using ECCs are not applicable or affordable on STT-MRAM tag array\cite{zhang2017shielding, takemura2010highly, wang2015selective, 15-EDCC-zhao2011design, na2016read, zhao2009high, mittal2017addressing, aliagha2019react, eli-date, seyedzadeh2016leveraging}.
$a)$ Read current reduction is limited to a value that the cells are still readable beside its adverse effect on read latency.
$b)$ Overwriting all tag ways in each cache access extremely increases the energy consumption and imposes significant performance penalty due to high write latency of STT-MRAM cells in read-restore scheme.
$c)$ Frequent data patterns required for data compression schemes cannot be found in tag array containing requested addresses.
$d)$ ECCs should be limited to correcting single errors because of high energy, area, and performance cost of providing larger correction capabilities in the tag part.

In contrast to the previous work, our suggested scheme decreases read disturbance in tag array by minimizing the number of reads from tag bits.  
As mentioned, some schemes targeted energy saving in traditional SRAM-based tags by reducing the number of tag accesses\cite{calder1996predictive,inoue1999way,sleiman2012embedded, zhang2005way,mallya2015way, kim2019segmented,peng2006low,min2004partial}.
While these schemes can reduce the number of reads from tag array, they mainly focused on tiny L1-caches and loose their efficiency when employed in large last-level caches.
Nevertheless, to provide a fair evaluation, we compare our proposed scheme with the best-performing scheme, i.e., way prediction, in this category in addition to comparing it with the conventional tag array configuration as the baseline.

\begin{figure*}[t]
				\centering
				\subfloat[]{\includegraphics[width=0.48\linewidth]{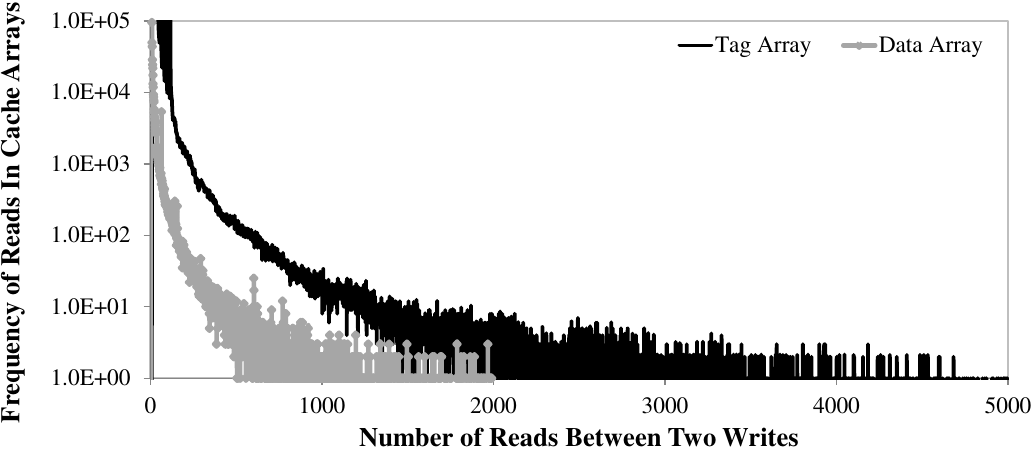}}
				\hspace{10pt}
				\subfloat[]{\includegraphics[width=0.48\linewidth]{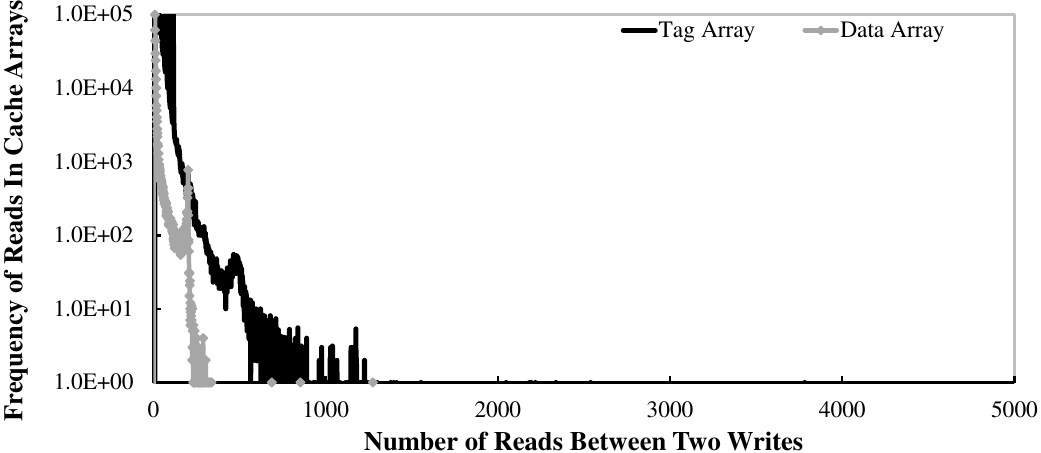}}
\hspace{10pt}
				\subfloat[]{\includegraphics[width=0.48\linewidth]{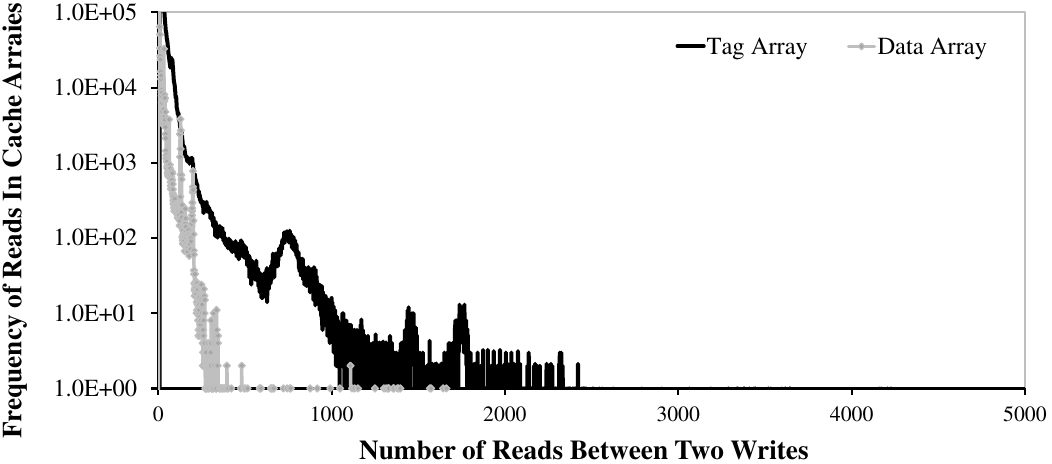}}
\hspace{10pt}
				\subfloat[]{\includegraphics[width=0.48\linewidth]{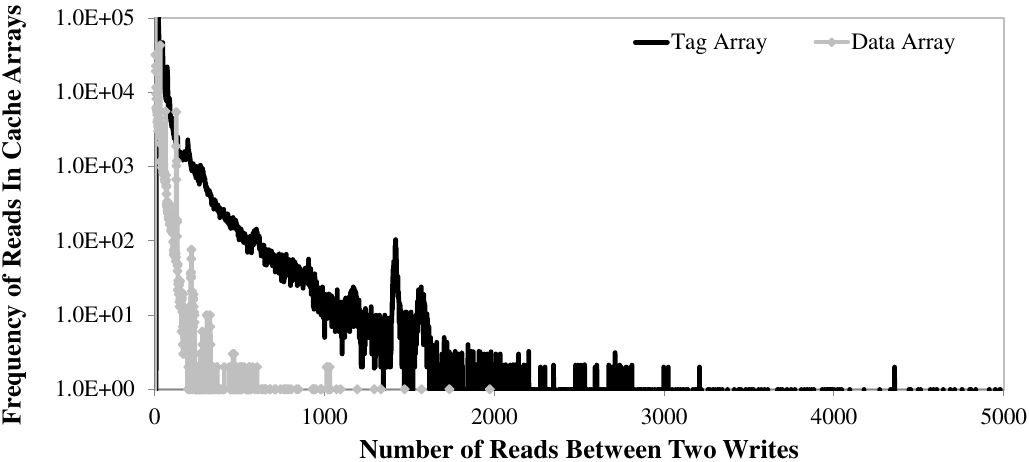}}
\hspace{10pt}
				\subfloat[]{\includegraphics[width=0.48\linewidth]{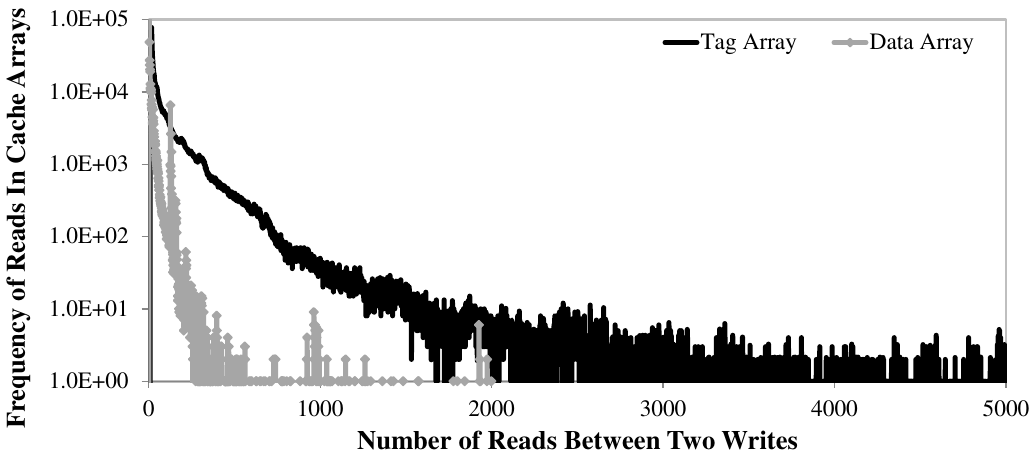}}
\hspace{10pt}
				\subfloat[]{\includegraphics[width=0.48\linewidth]{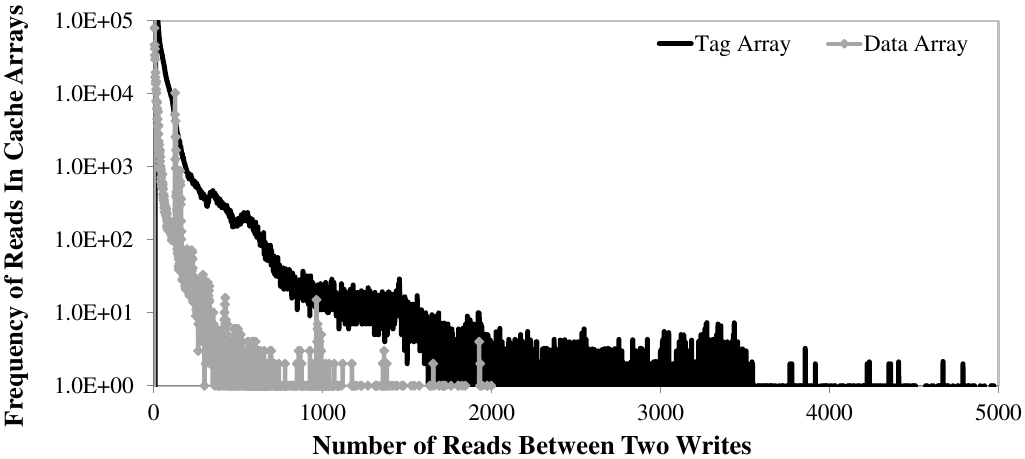}}

				\caption{Frequency of number of reads between two consecutive writes in data array vs tag array for (a) Mix0, (b) Mix3, (c) Mix4, (d) Mix5, (e) Mix6, and (f) Mix7 workloads.}\vspace{-10pt}
				\label{fig:3}
			\end{figure*}

\section{Proposed Scheme}

In this section, we first provide the problem definition and the source of the STT-MRAM cache unreliability, which is the motivation of this work. 
Then, the proposed 3RSeT scheme for reliability enhancement in the cache is explained.
Finally, the architectural details and implementation of 3RSeT are given.

\subsection{Problem Definition and Motivation} 
For an access request to a k-way set-associative cache, the tag part of the requested address is compared with all $k$ tags in the target set to find a match.
To minimize the tag operation time, all tags in the set are read in parallel and simultaneously compared with the incoming tag.
Therefore, $k$ tags are read for finding a single similar tag, if any.
Considering a 1MByte 8-way set-associative L2 cache with 64-byte block size in a system with 48-bit address bus, Fig.~\ref{fig:2} depicts the cache operation in an access request.
In this cache configuration, the number of sets is 2048 and the tag length is 31-bit.
When an access request is sent to the cache, the $index$ part of the incoming address is decoded in the first step. 
After determining the target set, all tags in the set are read and compared with the tag part of the address using the 31-bit comparators.

On a cache hit, one comparator finds a match and activates the corresponding way in data array to select the requested data block.
On a cache miss, on the other hand, none of the ways in data array is activated.
Reading all $k$ tags in a set to find a single similar tag, if any, significantly increases the number of read operations in the tag array.
In addition, both read and write requests to the cache require reading from the tag array.
Therefore, the number of reads from tag cells is substantially greater than that of from data cells.
The occurrence probability of read disturbance directly depends on the number of reads from STT-MRAM cells.
An error remains in the cell until the next write operation to the cell.
As a result, the number of read accesses to a STT-MRAM cell between two consecutive write operations determines the error probability.
Due to higher demand for read operation and less frequent write operation, the vulnerability of tag cells to the read disturbance is expected to be extremely higher than that of data cells.

We conduct a set of experiments to investigate the read access patterns and vulnerability of tag and data cells to read disturbance.
The gem5 full-system cycle-accurate simulator\cite{gem5} is used to model a quad-core processor and 18 combinations of programs from SPEC CPU2006 benchmark suite~\cite{spec2006} are generated as the multi-programmed  workloads.
The details of the system configuration and workloads are given in Section V (Simulation Setup and Results).

\begin{table}[t]
				\centering
				\caption{Read disturbance probability for a cell in data and tag arrays of STT-MRAM cache}
				\includegraphics[width=0.9\linewidth]{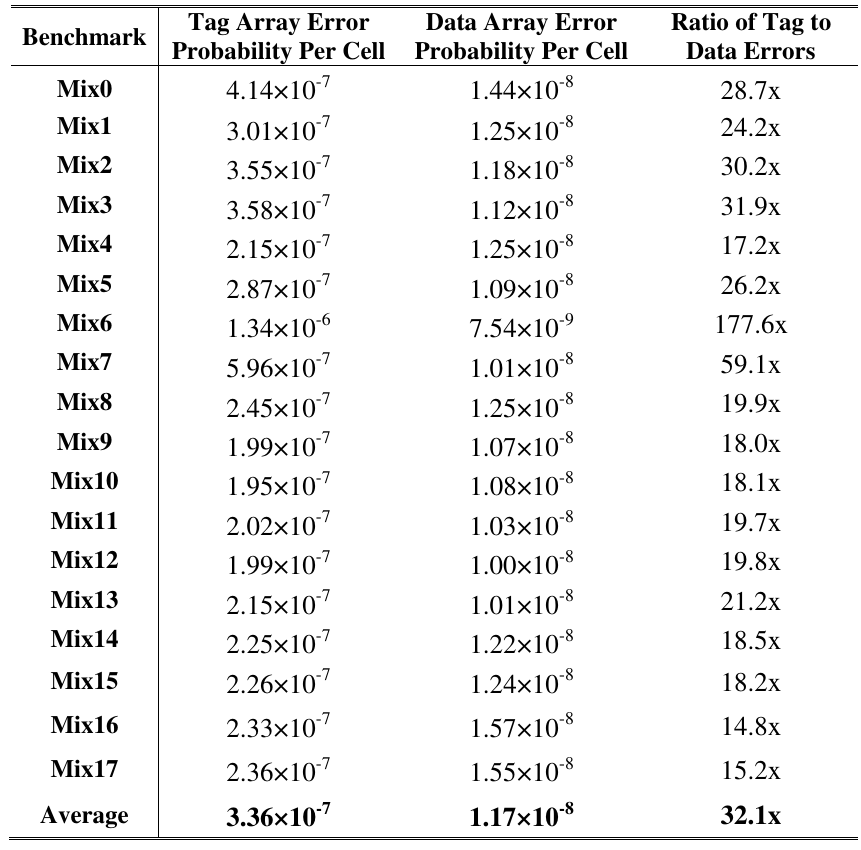}
				\label{table:1}\vspace{-10pt}
\end{table}

We extract the number of reads between all pairs of consecutive write operations in both data and tag arrays during the workload execution.
Fig.~\ref{fig:3} depicts the read access pattern in data and tag arrays for six exemplary workloads\footnote{The results for the other workloads are given in Appendix A.}.
The X-Axis shows the number of reads between two consecutive writes in linear scale and the Y-Axis in logarithmic scale shows the number of tags and data blocks that experienced each point in the X-Axis (the frequency of each number of reads).
For example, Point (1000, 100) for the tag array means that there are 100 tags during the workload execution that have been read for 1000 times between two consecutive writes.
The probability of read disturbance increases by moving from left to right on the X-Axis and from bottom to top on the Y-Axis.

According to Fig.~\ref{fig:3}, the frequency of tags and data decreases for larger number of reads between two consecutive writes.
However, the rate of this reduction is significantly higher for data blocks.
For small values of reads (e.g., 1 to 10) between two writes (left side of X-Axis), the frequency of tags is close to that of data blocks.
The data curve drops sharply and reaches close to zero at points larger than 200, where tag curve is still above 100.
As can be seen in Fig.~\ref{fig:3}-(a), Fig.~\ref{fig:3}-(e), and Fig.~\ref{fig:3}-(f), there is no data block with larger than 2000 reads, while several tags exist with near 5000 reads.

The maximum number of tag reads in Fig.~\ref{fig:3}-(b) is about 1300, which is by 5x larger than that of data block.
The curve for data blocks reaches close to zero at point 200 in Fig.~\ref{fig:3}-(c), while the curve for tag arrays remains near 10 until point 1000 with a peak to 100 in point 870.
The pattern observed between points 100 and 300 for the  curve of data blocks in Fig.~\ref{fig:3}-(d) before reaching it to zero is experienced in 2600 to 3000 interval for the curve of tag arrays. 
These observations indicate that in the case of both number of consecutive reads and the number of cells experiencing those reads, the values in the tag array are significantly greater than those in the data array.

To quantify the effect of the number of read operations on the read disturbance rate, we calculate the error probability in a STT-MRAM cell ($P_{Failure-per-cell}$) after $n$ consecutive read operations in (\ref{eq:5}).
\begin{flalign}
			\label{eq:5}
			\resizebox{.8\linewidth}{!}{$ P_{Failure-per-cell} = 1- {(1-P_{Read-Disturbance})^{n} }$}
			\end{flalign}
where, $n$ is the number of read operations and $P_{Read-Disturbance}$ is read disturbance probability of a cell for a single read operation given in (\ref{eq:1}).
Based on read access patterns extracted from our simulations and the results observed in Fig.~\ref{fig:3}, the read disturbance probability of a STT-MRAM in both tag and data arrays are calculated and given in Table~\ref{table:1} for all 18 workloads, i.e., $Mix0$-$Mix17$.
The read disturbance probability per read operation per cell is assumed to be $10^{-8}$~\cite{Eli-TC, naeimi2013intel,eli-date}.
As the results show, the error occurrence probability in a tag cell is by several times higher than that in a data cell for all workloads.
The tag error probability is by more than 50x higher than data error probability in $Mix6$ and $Mix7$. 
In the minimum gap, error probability in the tag is by 14.8x higher than that in the data for $Mix16$ workload.
On average, the read disturbance probability in tag array is by 32.1x higher than that in data array.
\textit{This observation confirms that the severity of read disturbance error in tag array is substantially higher than that in data array, even though all previous studies focused on the data array.}

The high vulnerability of STT-MRAM tag array to read disturbance error is because of the high demand for reading the tags.
The tag operation and configuration in STT-MRAM caches are inherited from traditional SRAM and DRAM caches, in which the read operation is not a reliability concern.
By restructuring the STT-MRAM cache architecture based on its characteristics, we can take advantages of this emerging technology and avoid its drawbacks due to its special opportunities and challenges.

\subsection{Proposed 3RSeT Scheme}

Our approach for mitigating read disturbance rate in tag array is to reduce the number of reads from tag cells per access request while maintaining its parallel comparisons feature.
To this aim, instead of comparing the entire tags, we selectively read and compare a small fraction of all tags and disable the large remaining fraction of the mismatched tags.
In the proposed scheme, so-called~$~\underline{R}ead$ \textit{Disturbance} $\underline{R}ate$ $\underline{R}eduction$ $by$ $\underline{Se}lective$ $\underline{T}ag$ $Comparison$ (3RSeT), the tag comparison operation is splitted into two steps.
First, a partial comparison is performed on $m$ lower order bits of the n-bit tags.
Then, the mismatched tag ways in partial comparison are disabled and the comparison is performed in the remaining $n-m$ bits of other tag ways.
By disabling some tag ways in the second step, their upper $n-m$ bits are immune to read disturbance.

The effectiveness of 3RSeT depends on the number of tags disabled in the second step.
Because of the locality in referring memory blocks and limited working set of applications, the chance for dissimilarity in lower order bits of tag is higher.
Therefore, it is expected that 3RSeT can effectively disable the majority of tags by partially comparing a few number of lower order bits.

\begin{figure*}[t]
				\centering
				\includegraphics[width=0.88\linewidth]{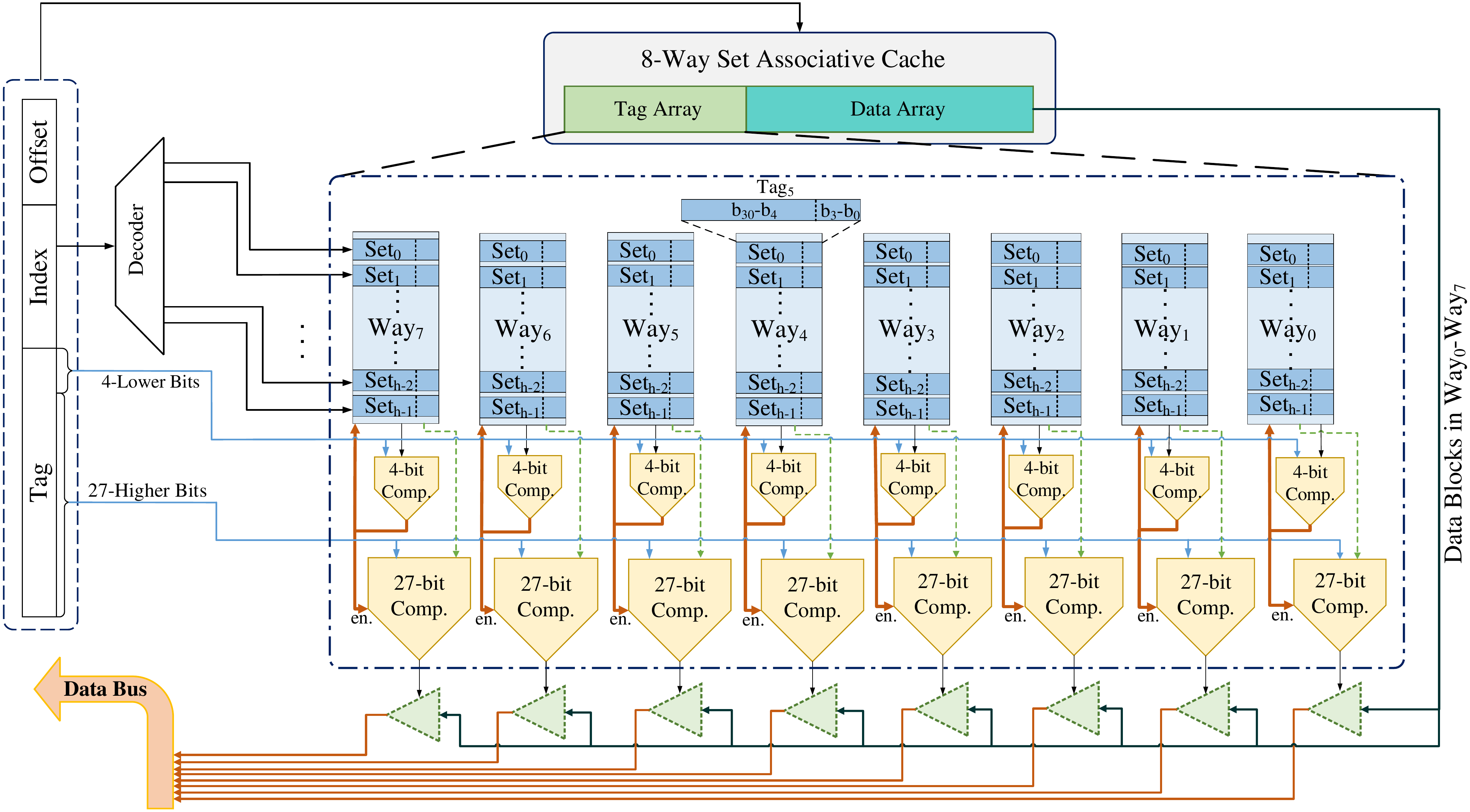}
				\caption{Structure of tag array in proposed 3RSeT scheme.}\vspace{-10pt}
				\label{fig:4}
\end{figure*}

\vspace{10pt}
\subsubsection{3RSeT Architectural Details}

Fig.~\ref{fig:4} illustrates the tag array configuration in 3RSeT.
We assume a 48-bit address bus in the system and the cache size is 1MByte~\cite{address64,intel-xeon, AMD64,intel64}.
The tag length for an 8-way set-associative cache is 31 bits and the cache block size is 64-Byte.
For the sake of simplicity, we limit our discussion to the mentioned parameters hereafter.
However, the explanations, discussions, and the proposed 3RSeT are generally valid and applicable.

3RSeT splits a 31-bit tag way into a 4-bit lower order part (bit$_{3}$-bit$_{0}$) and a 27-bit higher order part (bit$_{30}$-bit$_{4}$) (the reason behind selecting 4-bit as the lower order bits is discussed in the next subsection).
Considering the cache configuration in Fig.~\ref{fig:4}, the operation of tag array for a read access is as follows.
After decoding the $index$ part of the requested address, bit$_{3}$-bit$_{0}$ of all tags in the target set are read and compared with bit$_{20}$-bit$_{17}$ of the address (bit$_{3}$-bit$_{0}$ of the address tag part) via eight 4-bit comparators.
The output of each 4-bit comparator is connected to its corresponding tag way to determine whether the tag should be disabled for the next step.
In the next step, the tag ways that are similar to the input tag in their bit$_{3}$-bit$_{0}$ are activated while the other tags are disabled and the read operation is repeated.
Meanwhile, the output of 4-bit comparators are connected to their corresponding 27-bit comparator to disable those that are not required.
Disabling the comparator significantly saves the power in this step.
After reading bit$_{30}$-bit$_{4}$ of the active tag ways, their corresponding 27-bit comparators determine the target data block.

The key feature of 3RSeT is that the two step comparison is realized via a pure combinational circuit without requiring any modification in cache controller.
This internally-operated splitting minimizes the timing effect of 3RSeT on the tag comparison operation.
The implementation details will be explained in more depth in Section 4.2.3 and Section 5.3.

\vspace{10pt}
\subsubsection{Optimum Number of Low Order Bits}
3RSeT splits the tag way bits into two parts (high order bits and low order bits parts).
The first step of the 3RSeT scheme is to compare some lower order bits of the tag part of an incoming address with the corresponding lower order bits of all tags in the target set.  
The number of bits selected for comparison in this step can be any value between one to 30 for a 31-bit tag.
At the second step, 3RSeT compares the remaining bits of a subset of all tag ways with the remaining part of incoming tag.
On the other hand, dividing the tag comparison process into two steps needs to be completed in the predetermined time slot to impose no performance overhead.

To determine the partitioning border point in this 31-bit tag, we simulate the proposed 3RSeT for a wide range of values (one bit to ten bits) for partial tag length comparison in the first step.
In this simulation, one bit to ten bits out of 31-bit tag of all tags from the target set is compared with the incoming tag address at the first step.
In the second step, the remaining 30 to 21 bits of the matched tags are compared with the corresponding bits of the incoming tag address.
The number of bits that are read in each partitioning demonstrates the effectiveness of that partitioning.
The less number of bits read during an access, the higher effectiveness of that selection.

\begin{figure}[t]
				\centering
				\includegraphics[width=1\linewidth]{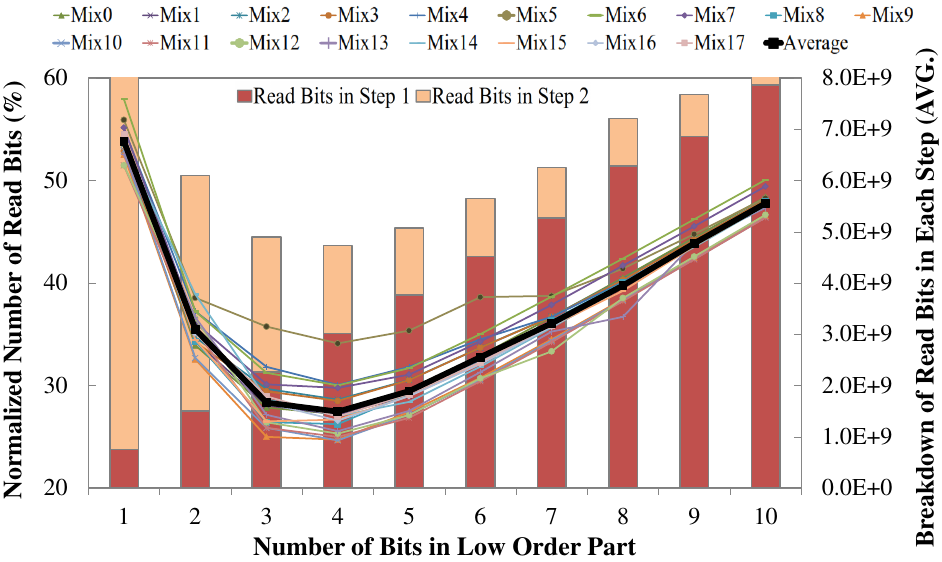}\vspace{-7pt}
				\caption{Number of reads from tag bits for different number of lower order bits to be compared in first step of 3RSeT.}\vspace{-8pt}
				\label{fig:5}
\end{figure}

Fig.~\ref{fig:5} illustrates the number of reads from tag bits for all accesses in 18 workloads from SPEC CPU2006 benchmark suite based on the tag splitted points. 
The X-Axis is the number of bits in 3RSeT compared in the first step of tag comparison.
The Y-Axis shows the total number of accessed bits in 3RSeT normalized to the conventional architecture, in which all the 31-bit tags are compared in one step.
As shown in Fig.~\ref{fig:5}, the minimum number of accessed bits for all workloads is observed at splitting point of four (comparison based on 4 lower order bits in the first step).
This observation reveals that the optimum length of lower order bits contributing in the first step is 4-bit.
Any value lower (higher) than 4-bit exacerbates the number of reads contributing in the second (first) step in such a way that leads to an increased number of total reads. 
The higher distance from 4-bit comparison in any direction, the larger number of total reads in the tag array.
Therefore, we choose four lower order bits as the first $m$ bits out of $n$ bits (31-bit tag) to compare in the first step of the 3RSeT scheme.  

%

\begin{figure}[t]
				\centering
				\includegraphics[width=1\linewidth]{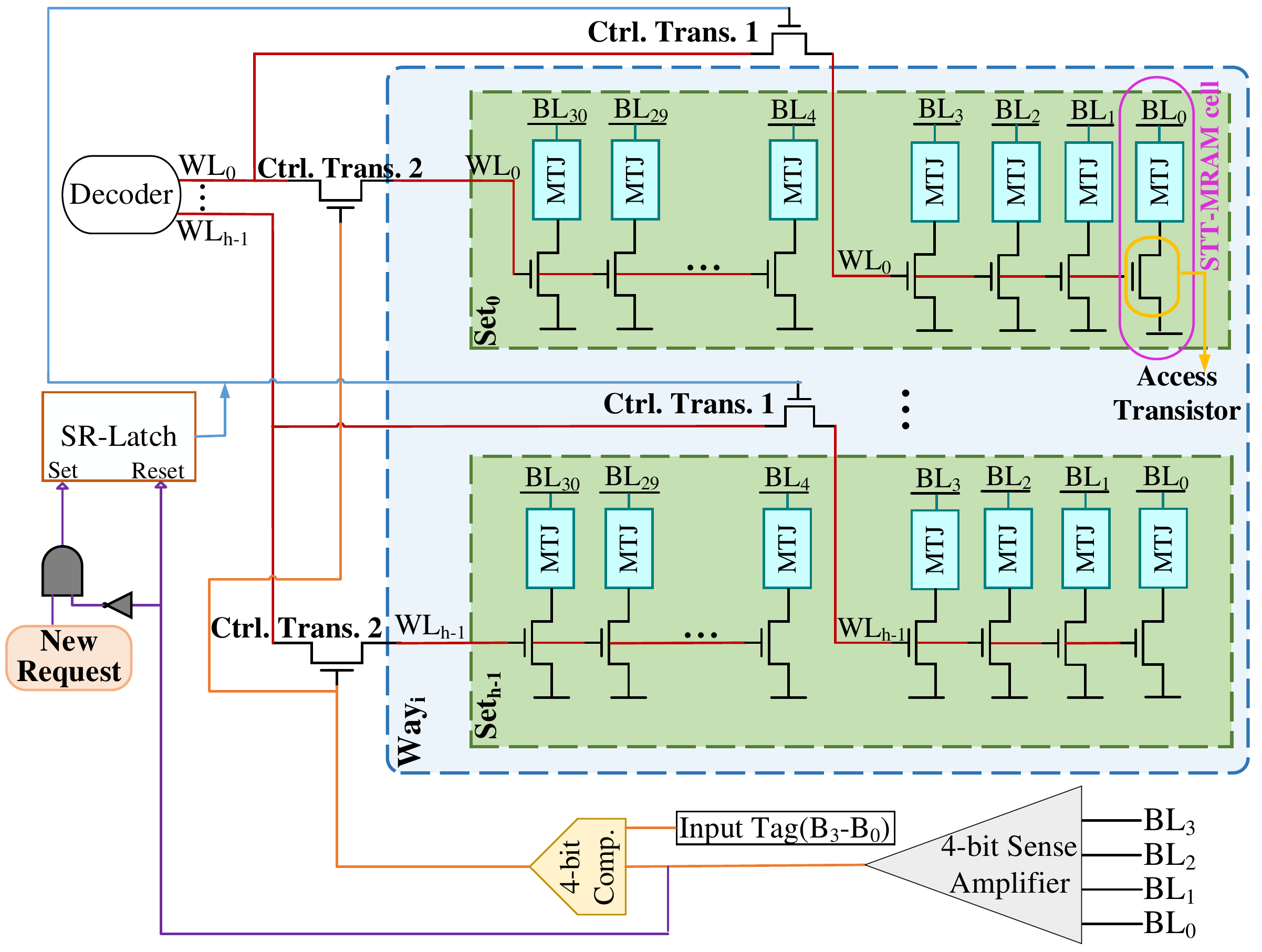}
				\caption{Internal organization of a tag way in proposed 3RSeT scheme.}\vspace{-10pt}
				\label{fig:6}
\end{figure}

\begin{figure*}[t]
				\centering
				\includegraphics[width=0.99\linewidth]{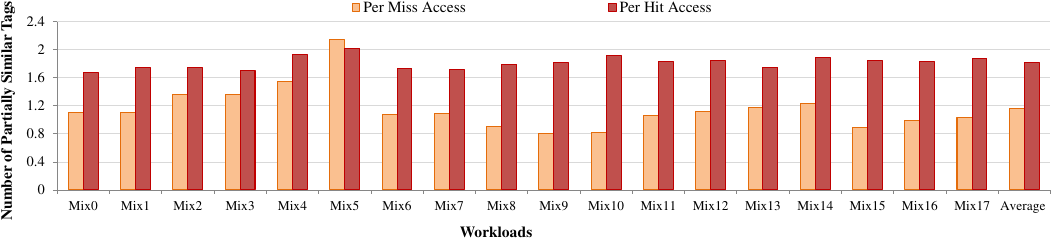}
				\caption{Average number of partially similar tags in the cache per hit and miss access per workload.}
				\label{fig:7}
\end{figure*}

\vspace{10pt}
\subsubsection{3RSeT Implementation Details}
To be able to partially read a tag line, i.e., to read bit$_{3}$-bit$_{0}$ in the first step and bit$_{30}$-bit$_{4}$ in the second step, the structure of tag ways should be modified.
3RSeT provides this capability with a minor modification in tag lines structure and a simple logic.
Fig.~\ref{fig:6} shows the detailed structure of a tag way in 3RSeT.
In the conventional structure, each output signal of $Decoder$ unit is directly connected to all 31 access transistors of 31-bit tag line.
This signal is the $Word$~$Line$ (WL) signal of the tag row.
In 3RSeT, the WL signal derived by the Decoder unit is connected to the access transistor through two control transistors, i.e., $Ctrl.~Trans.1$ and $Ctrl.~Trans.2$.
WL activated by decoder unit is connected to bit$_{3}$-bit$_{0}$ in each row through $Ctrl.~Trans.1$ and to bit$_{30}$-bit$_{4}$ through $Ctrl.~Trans.2$. 
The controlling logic to turn on/off the control transistors operates in such a way that in the first access step, the WL for  bit$_{3}$-bit$_{0}$ and  bit$_{30}$-bit$_{4}$ is enabled and disabled, respectively.
In the second step, WL for  bit$_{3}$-bit$_{0}$ is disabled while WL for  bit$_{30}$-bit$_{4}$ is enabled only for the tag ways that their 4-bit comparator signals a partial match.


The control transistor of  bit$_{30}$-bit$_{4}$ in each tag way, i.e., $Ctrl.~Trans.2$, is activated directly by the output of its corresponding 4-bit comparator.
To enable control transistor of bit$_{3}$-bit$_{0}$ in the first step and disable it in the second step, we use a $SR$-$Latch$ unit for each tag way.
On a new request, the SR-Latch is set by the rising edge of the request and activates the $Ctrl.~Trans.1$ to read from bit$_{3}$-bit$_{0}$.
The $Reset$ input of SR-Latch is triggered by the completion of 4-bit $Sense~Amplifier$ unit sensing bit$_{3}$-bit$_{0}$ and the $Set$ input is deactivated by the $AND$ gate. 
The output of SR-Latch deactivates the $Ctrl.~Trans.1$ in the second step, which remains inactivate until the next access request.

\section{Simulation Setup and Results}

We evaluate the proposed 3RSeT scheme using gem5 full-system cycle-accurate simulator~\cite{gem5}.
A quad-core processor with private L1 instruction- and data-cache and an L2 cache shared between the cores is modeled.
The details of the system configuration is shown in Table~\ref{table:3}.
We implement 3RSeT in the L2 cache assuming that 31-bit tags are splitted into a 4-bit lower and 27-bit upper parts. 

In the experiments, 18 multi-programmed workloads are generated by combining a set of programs from SPEC CPU2006 benchmark suite~\cite{spec2006}.
The results are extracted by executing 4 billion instructions after skipping initial 400 million instructions as the warm-up phase.
The probability of read disturbance occurrence per read access to a single STT-MRAM cell is $10^{-8}$ in the experiments~\cite{naeimi2013intel, Eli-TR}. 
3RSeT is compared with the STT-MRAM conventional cache that the configuration is mentioned in Table~\ref{table:3} as $baseline$ and $way~prediction$, as a well-known read reduction scheme, in terms of the number of tag bits read and error rate as well as tag energy consumption, performance, and area.
As discussed in Section 3, several schemes have been presented in the literature to reduce the number of reads from cache.
Almost all schemes presented for read reduction in the cache focused on data array and only way prediction scheme is capable of tag bits read reduction.
Although all of these schemes targeted L1 cache, investigating way prediction reveals the deficiency of the previous schemes in L2 caches.

\begin{table}[t]
				\centering
				\caption{System configuration details}
				\includegraphics[width=1\linewidth]{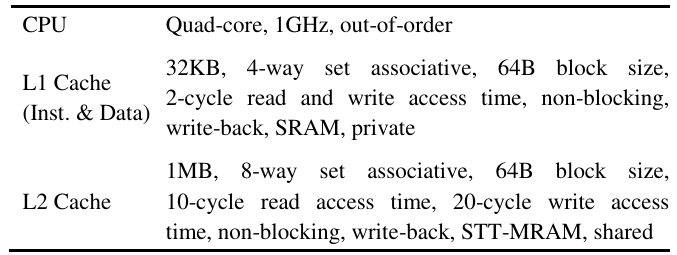}
				\label{table:3}\vspace{-10pt}
\end{table}

\subsection{Reliability Evaluation}
3RSeT improves the tag reliability by mitigating the read disturbance error rate, which is achieved by reducing the number of read accesses to tag cells.
This reduction is realized by eliminating read operation from upper 27-bit part of those tags that their lower 4-bit part is dissimilar to that of input tag.
Fig.~\ref{fig:7} depicts the average number of similar tags for both cache hits and cache misses in each workload.
The results illustrate that, on average, only 1.88 out of 8 tags are partially similar to input tags in cache hits.
This average value is as low as 1.03 for cache misses.
In the worst-case, an average of 2.14 out of 8 tags are partially similar in $Mix5$ workload.
This observation reveals that 3RSeT effectively disactivates more than 73\% of tag ways per access.

\begin{figure*}[t]
				\centering
				\includegraphics[width=0.99\linewidth]{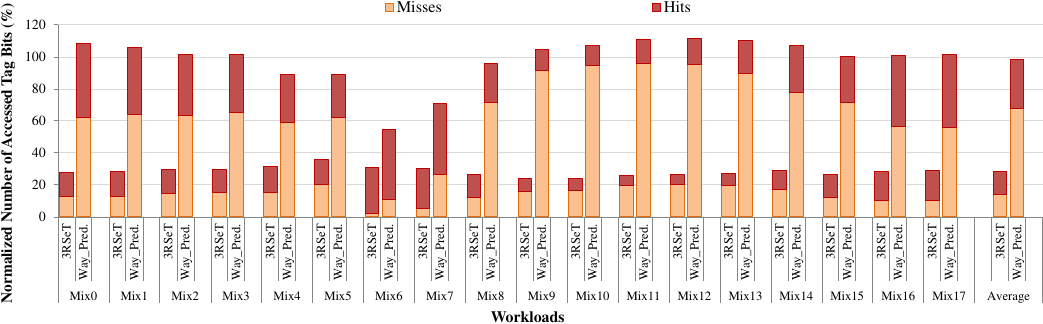}\vspace{-4pt}
				\caption{Breakdown of total number of accessed tag bits in hit and miss requests for 3RSeT and way prediction (normalized to the baseline for all workloads).}\vspace{-7pt}
				\label{fig:8}
\end{figure*}

Fig.~\ref{fig:8} depicts the total number of accessed tag bits in 3RSeT and way prediction schemes normalized to the baseline.
The contribution of cache hits and cache misses is illustrated for each workload.
On average, the number of bits accessed using 3RSeT and way prediction is 28.2\% and 98.3\%, respectively, indicating the reduction of 71.8\% in the former and less than 2\% in the latter.
Read reduction in 3RSeT is between 65\% to 70\% in only three workloads, i.e., $Mix4$-$Mix6$, and is more than 75\% in $Mix9$-$Mix11$.

\begin{figure*}[b]
				\centering\vspace{-5pt}
				\includegraphics[width=1\linewidth]{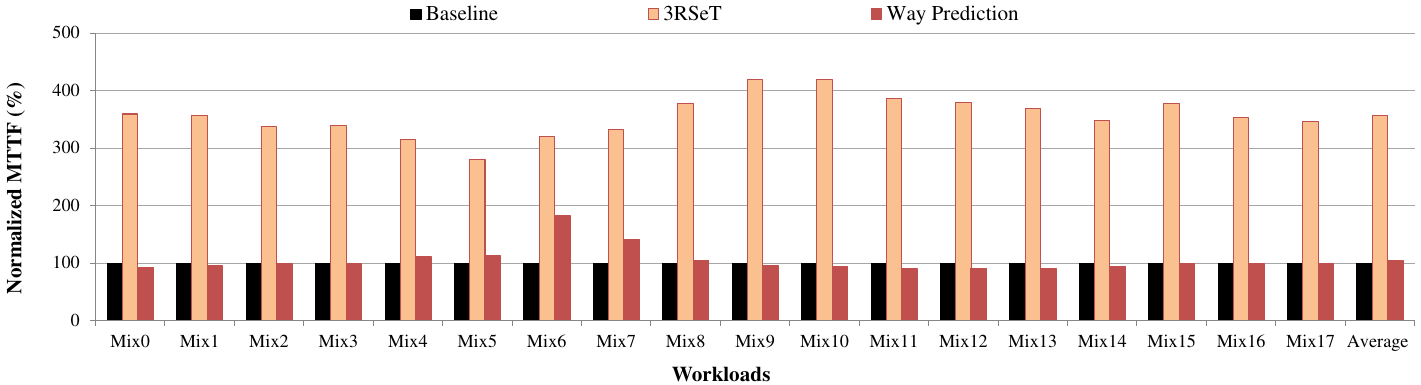}\vspace{-5pt}
				\caption{Mean Time To Failure (MTTF) of tag array in 3RSeT and way prediction normalized to the baseline for all workloads.}\vspace{-5pt}
				\label{fig:9}
\end{figure*}

Way prediction scheme reduces the number of tag bit reads only for five workloads, i.e., $Mix4$-$Mix8$, whereas, this number is increased for all other workloads up to 111.2\% compared to the baseline.
This increase in comparison with baseline is due to mispredictions in the way prediction scheme.
On a correct prediction, one out of eight tag ways  (in an 8-way set-associative cache) is read, while eight tag ways in addition to the one predicted tag are read on a misprediction.
Misprediction is experienced in all cache misses and in a subset of cache hits.
In this regard, a large fraction of total number of reads in the way prediction scheme is devoted to the cache misses, as observed in Fig.~\ref{fig:8}. 
Cache misses contribute by 67.3\%, on average, and by more than 90\% in $Mix9$-$Mix13$ workloads in total number of reads.

To explore the effect of the number of reads on the tag reliability, we compare the $Mean$~$Time$~$To$~$Failure$~(MTTF) parameter of the 3RSeT and way prediction schemes.
MTTF is extracted from the error rate per unit of time calculated based on error probability per tag access.
Fig.~\ref{fig:9} reports the MTTF in 3RSeT and way prediction normalized to the baseline for all workloads.
The average of MTTF in 3RSeT and way prediction is 357.9\% and 104.9\%, respectively.
MTTF in 3RSeT is even higher than 420\% in $Mix9$ and $Mix10$ workloads and is slightly less than 300\% in only $Mix5$ workload.
This value in the way prediction scheme is lower than the baseline for several workloads, e.g., $Mix0$-$Mix3$ and $Mix9$-$Mix17$, and its maximum value is 183.3\% for $Mix6$ workload. 
Therefore, 3RSeT extends MTTF to 3.6x, while way prediction achieves a negligible MTTF improvement of 1.05x.

\begin{figure}[t]
				\centering\vspace{-5pt}
				\includegraphics[width=0.99\linewidth]{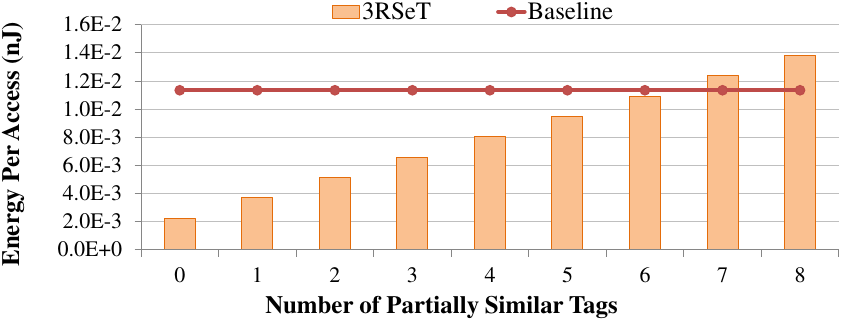}
				\caption{Energy Consumption of tag array per access in conventional cache configuration (baseline) and 3RSeT for all nine possible scenarios of tags partial similarity.}\vspace{-15pt}
				\label{fig:10}
\end{figure}

\begin{figure*}[t]
				\centering\vspace{-10pt}
				\includegraphics[width=0.99\linewidth]{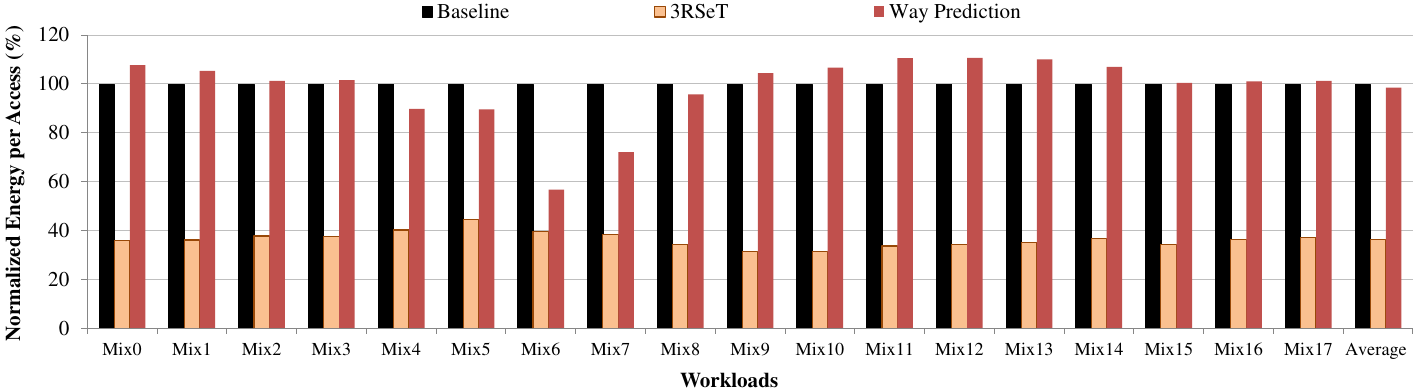}\vspace{-6pt}
				\caption{Energy consumption of tag array in 3RSeT and way prediction normalized to the baseline for all workloads.}
				\label{fig:11}
\end{figure*}

\subsection{Energy Consumption Evaluation}

By modifying the tag array organization and reducing the number of reads from tag cells, 3RSeT affects the energy consumption in tag array per access request.
On the one hand, reading and comparing the lower and higher tag parts in two steps slightly increases the energy consumption related to that tag way.
On the other hand, eliminating the read and comparison operation of the higher tag part reduces the energy consumption in tag ways disabled in the second step of 3RSeT.
Note 3RSeT reduces energy consumption in the tag array if its energy saving due to disabling some tag ways outperforms the energy consumption increase in tag ways that remain active in the second comparison step of 3RSeT.
Therefore, the energy efficiency of 3RSeT per access depends on the number of disabled tag ways.

Fig.~\ref{fig:10} depicts the tag array energy consumption in the baseline, in which all eight tag ways are entirely read and compared with input tag, and in 3RSeT for all nine possible scenarios in term of the partial similarity.
These nine scenarios for an 8-way set-associative cache is the similarity of zero, one, ..., seven, or all eight tag ways with the incoming tag.
In the worst-case scenario, lower order bits of all eight tags are similar to the input tag and no tag way is disabled in the second step.
As shown, the energy consumption in the baseline is 0.01nJ, which is slightly lower than that in 3RSeT when the lower part of seven tags are similar to input tag.
In this case, only one of the tag ways is disabled in the second comparison step and 3RSeT provides minimum reduction in the number of bits read.
For all other seven scenarios, energy consumption in 3RSeT is lower than that of the baseline.
The tag energy consumption in 3RSeT is 45.4\% and 32.8\% of the baseline when two and one partial similar tags are found, respectively.
This value is 19.9\% for the case that no partial similarity exists.

Fig.~\ref{fig:11} shows the tag array energy consumption in the 3RSeT and way prediction schemes normalized to the baseline for all workloads.
Way prediction reduces the energy consumption by only 1.6\%, on average.
This scheme reads and compares a single predicted tag way on a correct prediction, which results in significant energy saving.
On the other hand, one predicted tag way and then all the eight tag ways are read and compared on a misprediciton case, which increases the energy consumption in comparison to the baseline.
The overall effect of way prediction on the energy consumption depends on its prediction accuracy.
Based on the results, it is obvious that the misprediction rate is as high that penalize the energy saving of the correct predictions.
This penalty for some workloads, e.g., $Mix0$-$Mix3$, and $Mix9$-$Mix14$, is even so high that the total energy consumption in way prediction is larger than the baseline.
In contrary to way prediction, 3RSeT effectively reduces the energy consumption by 62.1\%, on average.
This reduction varies from 53.4\% in $Mix5$ workload to 67.4\% in $Mix9$ workload.
The trend in tag energy consumption by 3RSeT is according to the reduction in the number of tag reads observed in Fig.~\ref{fig:8}.
The lower number of accessed tag cells per read, the higher energy saving is achieved by 3RSeT.

\subsection{Performance Evaluation}
We investigate the effects of the 3RSeT and way prediction schemes on the STT-MRAM cache performance.
Performance overhead of 3RSeT depends on how it affects the cache access time.
Table~\ref{table:performance} shows cache access time in the baseline and 3RSeT for various cache size and associativities.
The cache access time consists of two parts, i.e., tag array and data array.
Since data array and tag array are accessed simultaneously, the total cache access time is the maximum of the two.
3RSeT imposes no performance penalty if after its modifications, the output of the comparators are still ready before arriving the data blocks, assuming this is the case for conventional cache configuration.

As depicted in Table~\ref{table:performance}, delay of the data array is largely higher than the tag array for all cache configurations in the baseline.
3RSeT splits the 31-bit tag comparison to the 4-bit comparison followed by a 27-bit comparison.
This two-step sequential comparison operation conducted by a combinational controlling circuit in single clock cycle increases the tag array delay in 3RSeT.
According to the results in Table~\ref{table:performance}, the tag array delay after 3RSeT modifications is still lower than that of the data array.
Therefore, cache access time remains intact in a 3RSeT-equipped tag array.

\begin{table}[t]
				\centering
				\caption{Timing parameters for different cache sizes in 8- and 16-way set-associative configurations.}
				\includegraphics[width=1\linewidth]{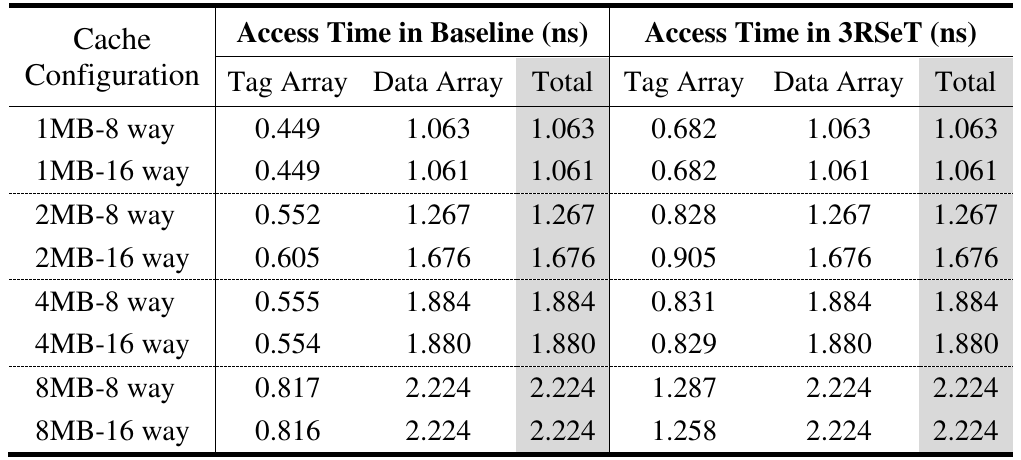}
				\label{table:performance}\vspace{-15pt}
\end{table}

\begin{figure*}[t]
				\centering
				\includegraphics[width=1\linewidth]{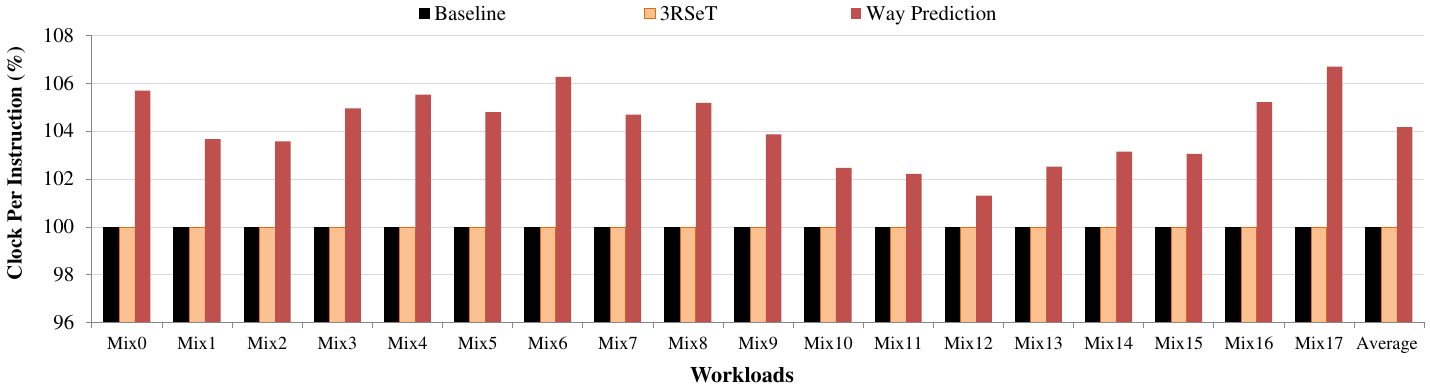}\vspace{-6pt}
				\caption{Clock Per Instruction (CPI) in 3RSeT and way prediction normalized to the baseline for all workloads.}\vspace{-5pt}
				\label{fig:12}
\end{figure*}

Fig.~\ref{fig:12} shows $Cycle~Per~Instruction$ (CPI) of 3RSeT and way prediction normalized to the baseline for all workloads.
The results show that way prediction degrades the performance by 4.2\%, on average.
This degradation in the worst-case is 6.7\% for $Mix17$ workload and its minimum value is 1.3\% in $Mix12$ workload.
The performance overhead in way prediction depends on its prediction accuracy. 
While the cache access time on a correct prediction is the same as that in the baseline, the operations of tag and data arrays on a mispredition need to be repeated for all cache ways.
Therefore, a higher delay is experienced on a mispredition, resulting into performance degradation.

\subsection{Area Evaluation}
The last parameter evaluated in this section is the cache area in 3RSeT.
Our implementations illustrate that the area overhead imposed by 3RSeT is negligible.
As shown in Fig.~\ref{fig:4}, 3RSeT architecture transforms a 31-bit comparator of the baseline into a 4-bit and a 27-bit comparator for each way.    
The overall area of a 4-bit and a 27-bit comparators is almost equal to the area occupied by a 31-bit comparator.
Meanwhile, the area of the comparators is negligible in total cache area.

The other modification in 3RSeT was shown in Fig.~\ref{fig:6}.
Each tag way is equipped with a $SR$-$Latch$, an $AND$ gate, and a $NOT$ gate, which results in including eight units of these modules to the total cache structure.
On the other hand, 3RSeT adds two NMOS transistors to each cache line according to Fig.~\ref{fig:6}.
Considering 1-MByte STT-MRAM cache, the area overhead of the added SR-Latches, AND gate, NOT gate, and NMOS transistors is less than 0.4\%.
It is noteworthy that the storage added for keeping the prediction information in way prediction imposes about 0.1\% area overhead.

\section{Limitations and Discussion}
This section discusses 3RSeT from various aspects and explores its limitations.
As investigated in Section 4.1, tag array in STT-MRAM caches is significantly more vulnerable to read disturbance than data array.
However, the error rate in data array cannot be ignored and a reliable STT-MRAM cache requires protecting both data and tag arrays.
The proposed 3RSeT scheme takes advantages of memory reference locality to reduce the number of reads from tag array.
However, this locality cannot be exploited in the data array, which makes 3RSeT inapplicable in this cache part.
As mentioned in Section 3, several schemes have been presented in the literature to protect data array against read disturbance, e.g., ECCs and REAP-Cache\cite{eli-date,naeimi2013intel, seyedzadeh2016leveraging, farbeh2016floating}.
Our proposed scheme is compatible with the previous schemes and can be jointly used with them to protect both tag and data arrays.

From performance perspective, 3RSeT increases the tag operation delay by splitting the tag comparison into two steps.
If not well managed, these two-phased operations may result in increasing the total cache access time.
However, as described in Section 4.2.3, we propose a pure combinational controlling circuitry that minimizes the delay effect of 3RSeT on tag operation delay; and, as demonstrated in Section 5.3, 3RSet opportunistically exploits the gap between delay of data and tag arrays operations to completely mask the increase in tag array delay.
Therefore, the cache access time is $not$ affected in 3RSeT despite the increased delay of tag operation.

From another perspective, workloads behavior and cache configuration may affect the 3RSeT efficiency and improvements.
As depicted in Fig.~\ref{fig:7}, the reduction in the number of tag reads by 3RSeT for cache misses is higher than cache hits.
Workloads with higher randomness in accessing memory blocks experience higher cache miss rate and more read reduction. 

On the other hand, miss rate typically increases in smaller caches, resulting in higher 3RSeT improvement by shrinking the cache size.
The tag length affects 3RSeT efficiency, as well.
While the optimum number of bits compared in the first step of 3RSeT is four regardless of the tag length, a greater number of bits is disabled in the second step for wider tags.
For a fixed address bus width, the tag length is wider in smaller caches. 
Tag length is also affected by the cache associativity.
The number of cache sets is reduced in higher associativities, which again leads to wider tags.
Meanwhile, tag length is directly determined by the address bus width.

Lastly, our evaluations in this work are based on a 48-bit address bus.
This value is the typical address bus width in modern $Intel$ processors~\cite{intel64, intel-xeon}.
In general, address bus width varies from 40-bit to 64-bit in different processor series and various vendors.
For example, while address bus in $ARM~Cortex$-$A35$ and $Cortex$-$A55$ are 40-bit, $ARM~Cortex$-$A57$ supports physical address space of both 40- and 44-bit~\cite{arm35, arm55, arm57}.
A more recent ARM product, i.e., $Cortex$-$A77$, is designed based on 52-bit address width~\cite{arm77}.
Both $Intel64$ and $AMD64$ processor families potentially support address width of 64-bit and are fabricated nowadays for 48-bit~\cite{intel64,AMD64,  address64, intel-xeon}.
While 3RSeT is applicable to all these architectures, significantly greater energy saving and error rate reduction is expected to achieve in processors with wider address width. 

\section{Summary and Conclusions}
Conventional on-chip cache structures lose their efficiency by replacing the prevalent SRAM memory cells with the emerging STT-MRAM technology.
This paper first revealed that excessive read operations in tag array for parallel comparison severely degrade the cache reliability, which is due to vulnerability of STT-MRAM cells to read disturbance.
Then, we proposed the 3RSeT scheme to redesign the tag array configuration according to the STT-MRAM reliability requirements.
Compared to the conventional configuration, 3RSeT provides 3.6x higher MTTF for tag array by eliminating a large fraction of unnecessary reads.
In addition, the proposed scheme reduces the energy consumption in the tag array by 62.1\% without compromising the cache performance.
This paper is a pioneer effort to investigate the efficiency of conventional cache configurations and redesign the cache according to new challenges, opportunities, and characteristics of STT-MRAM. 
			\vspace{-5pt}
			 \section*{Acknowledgments}
			%
			%
			This work has been supported in part by $Iran~National~Science~Foundation$ (INSF) under grant number 96006071 and in part by $Iran~National~Elites~Foundation$.
			
			\ifCLASSOPTIONcaptionsoff
			\newpage
			\fi

			
			
			\bibliographystyle{IEEEtran}
			\bibliography{references}
			\begin{IEEEbiography}[{\includegraphics[width=1in,height=1.25in,clip,keepaspectratio]{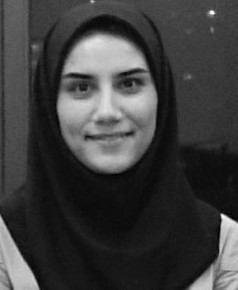}}]{Elham Cheshmikhani}
				received the B.Sc. degree in computer engineering from Iran University of Science and Technology (IUST) and the M.Sc. degree in computer engineering from Amirkabir University of Technology (Tehran Polytechnic-AUT), Tehran, Iran, in 2011 and 2013, respectively. She is currently a PhD candidate in computer engineering at Sharif University of Technology (SUT), Tehran, Iran.
She was a member of Design and Analysis of Dependable Systems (DADS) at AUT from 2011 to 2015, and has been a member of the Dependable Systems Laboratory (DSL) and Data Storage, Networks \& Processing Laboratory (DSN) since 2015 and 2017, respectively. Her research interests include emerging nonvolatile memory technologies, dependability analysis, fault tolerance, and storage systems. 
More recently, she received the Best Paper Award at IEEE/ACM Design, Automation, and Test in Europe (DATE) in 2019.
		\end{IEEEbiography}

%
%
%

			\vspace{-20pt}

			\begin{IEEEbiography}[{\includegraphics[width=1in,height=1.25in,clip,keepaspectratio]{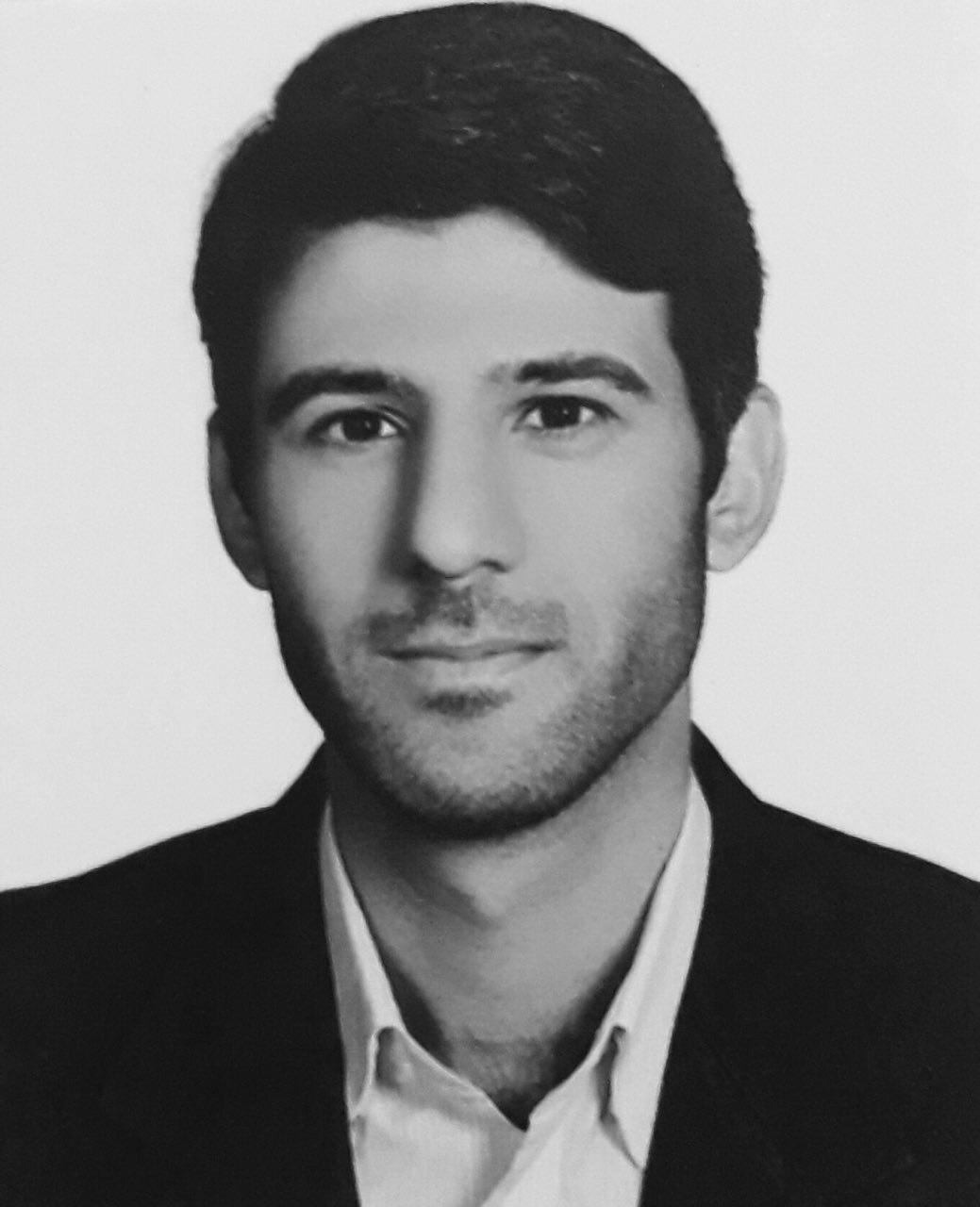}}]{Hamed Farbeh}
(S'12-M'19) received the B.Sc., M.Sc., and PhD degrees in computer engineering from Sharif University of Technology (SUT), Tehran, Iran, in 2009, 2011, and 2017, respectively.
He was a member of the Dependable Systems Laboratory (DSL) at SUT from 2007 to 2017 and the head of DSL from April 2017 to February 2018. 
He is currently the faculty member of the Department of Computer Engineering, Amirkabir University of Technology (Tehran Polytechnic-AUT), Tehran, Iran, where he established Intelligent Computing and Communication Infrastructure Laboratory (ICCI). 
He is also the head of the Internet-of-Things (IoT) research center at AUT and the member of the board of Cyber-Physical Systems Society of Iran (CPSSI).
He was with the Embedded Computing Laboratory (ECL), KAIST, Daejeon, South Korea, as a Visiting Researcher from October 2014 to May 2015 and collaborated with the Institute of Research for Fundamental Sciences (IPM), Tehran, Iran, as Postdoc fellow from May 2017 to January 2018.
He received the Best Paper Award at the IEEE/ACM Design, Automation, and Test in Europe conference (DATE) in 2019 for his work on STT-MRAM cache memory.
His current research interests include reliable memory hierarchy, emerging memory technologies, AI processors, and cyber-physical systems.

			\end{IEEEbiography}
	\vspace{-20pt}
			\begin{IEEEbiography}[{\includegraphics[width=1in,height=1.25in,clip,keepaspectratio]{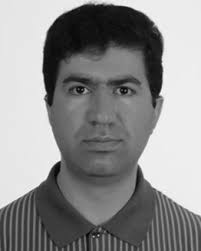}}]{Hossein Asadi}
				(M'08, SM'14) received the B.Sc. and M.Sc. degrees in computer engineering from the SUT, Tehran, Iran, in 2000 and 2002, respectively, and the Ph.D. degree in electrical and computer engineering from Northeastern University, Boston, MA, USA, in 2007. 
He was with EMC Corporation, Hopkinton, MA, USA, as a Research Scientist and Senior Hardware Engineer, from 2006 to 2009. From 2002 to 2003, he was a member of the Dependable Systems Laboratory, SUT, where he researched hardware verification techniques. From 2001 to 2002, he was a member of the Sharif Rescue Robots Group. He has been with the Department of Computer Engineering, SUT, since 2009, where he is currently a tenured Associate Professor. He is the Founder and Director of the \emph{Data Storage, Networks, and Processing} (DSN) Laboratory, Director of Sharif \emph{High-Performance Computing} (HPC) Center, and the President of Sharif ICT Innovation Center. He spent three months in the summer 2015 as a Visiting Professor at the School of Computer and Communication Sciences at the Ecole Poly-technique Federele de Lausanne (EPFL). He is also the co-founder of HPDS corp., designing and fabricating midrange and high-end data storage systems. He has authored and co-authored more than eighty technical papers in reputed journals and conference proceedings. His current research interests include data storage systems and networks, solid-state drives, operating system support for I/O and memory management, and reconfigurable and dependable computing.
Dr. Asadi was a recipient of the Technical Award for the Best Robot Design from the International RoboCup Rescue Competition, organized by AAAI and RoboCup, a recipient of Best Paper Award at the 15th CSI International Symposium on \emph{Computer Architecture \& Digital Systems} (CADS), the Distinguished Lecturer Award from SUT in 2010, the Distinguished Researcher Award and the Distinguished Research Institute Award from SUT in 2016, and the Distinguished Technology Award from SUT in 2017. He is also recipient of Extraordinary Ability in Science visa from US Citizenship and Immigration Services in 2008. He has been ranked among "Top-10" among 500+ faculties by Research and Technology Deputy, Sharif University of Technology for three consecutive years from 2016 to 2018. More recently, he received the Best Paper Award at IEEE/ACM Design, Automation, and Test in Europe (DATE) in 2019.
He has served as a Guest Editor of IEEE Transactions on Computers, an Associate Editor of Microelectronics Reliability, a Program Co-Chair of CADS2015, and the Program Chair of CSI National Computer Conference (CSICC2017).

			\end{IEEEbiography}

			\end{document}